\documentclass[10pt,journal,twoside]{IEEEtran}

\usepackage{multirow}
\usepackage{amssymb}
\usepackage[dvips]{graphicx}
\usepackage{amsmath}
\usepackage{amsthm}
\usepackage{latexsym,bm}
\usepackage{color}
\usepackage{subfigure}
\usepackage{longtable}
\usepackage{accents}
\usepackage{cite}
\usepackage{enumerate}
\usepackage{arydshln}
\usepackage{slashbox}
\usepackage{algorithm}
\usepackage{algorithmic}
\usepackage{float}
\usepackage{booktabs}
\usepackage{subfigure}
\usepackage{stfloats}
\usepackage{graphicx}
\usepackage{bm}
\usepackage{mathrsfs}
\usepackage{ragged2e}

\makeatletter
\def\widebar{\accentset{{\cc@style\underline{\mskip10mu}}}}
\def\Widebar{\accentset{{\cc@style\underline{\mskip8mu}}}}
\makeatother

\theoremstyle{plain}

\theoremstyle{definition}
\theoremstyle{definition}

\usepackage{lineno}

\begin{document}	
\title{\huge Affine Frequency Division Multiplexing with Index Modulation: Full Diversity Condition, Performance Analysis, and Low-Complexity Detection
\thanks{The work of M. Wen was supported in part by the National Key R$\&$D Program of China under Grant 2023YFB2904500, and in part by the Fundamental Research Funds for the Central Universities under Grant 2024ZYGXZR076.
The work of Y. Ge was supported in part by the RIE2020 Industry Alignment Fund-Industry Collaboration Projects (IAF-ICP) Funding Initiative, as well as cash and in-kind contribution from the industry partner(s).
The work of J. Li was supported in part by Guangdong Basic and Applied Basic Research Foundation under Grant 2023A1515030118; the Guangzhou Science and Technology Project under Grant 2023A03J0110; the Key Discipline Project of Guangzhou Education Bureau under Grant 202255467.
The work of N. Al-Dhahir was supported by Erik Jonsson Distinguished Professorship at UT-Dallas. {\em (Corresponding author: Miaowen Wen.)}}
\thanks{Yiwei~Tao, and Miaowen~Wen are with the School of Electronic and Information Engineering, South China University of Technology, Guangzhou 510641, China (e-mail: eeyiweitao@mail.scut.edu.cn; eemwwen@scut.edu.cn).}
\thanks{Yao~Ge is with the Continental-NTU Corporate Lab, Nanyang Technological University, 639798, Singapore (e-mail: yao.ge@ntu.edu.sg).}
\thanks{Jun~Li is with the School of Electronics and Communication Engineering, Guangzhou University, Guangzhou 510006, China (e-mail: lijun52018@gzhu.edu.cn).}
\thanks{Ertugrul Basar is with the Communications Research and Innovation Lab-oratory (CoreLab), Department of Electrical and Electronics Engineering, Ko\c{c} University, Sariyer 34450, Istanbul, Turkey. (e-mail: ebasar@ku.edu.tr).}
\thanks{Naofal Al-Dhahir is with the Department of Electrical and Computer Engineering, The University
of Texas at Dallas, Richardson, TX 75080 USA (e-mail: aldhahir@utdallas.edu).}
}
\author{\fontsize{11pt}{\baselineskip}\selectfont{
Yiwei Tao, \textit{Graduate Student Member, IEEE}, Miaowen~Wen, \textit{Senior Member, IEEE}, Yao~Ge, \textit{Member, IEEE},\\ Jun~Li, \textit{Senior Member, IEEE}, Ertugrul~Basar, \textit{Fellow, IEEE}, and Naofal Al-Dhahir, \textit{Fellow, IEEE}}
\vspace{-0mm}
}
\maketitle
\thispagestyle{empty}
\pagestyle{empty}
\begin{abstract}
Affine frequency division multiplexing (AFDM) is a novel modulation technique based on chirp signals that has been recently proposed as an effective solution for highly reliable communications in high-mobility scenarios.
In this paper, we focus on the design of robust index modulation (IM) schemes under the multiple-antenna AFDM transmission framework.
To this end, the cyclic delay diversity (CDD) technique is employed to harvest the transmit diversity gain.
As a result, we propose two novel AFDM-IM schemes with transmit diversity, termed as {\em CDD-AFDM-IM-I} and {\em CDD-AFDM-IM-II}.
We analyze the full diversity conditions and parameter settings of the proposed CDD-AFDM-IM schemes for both integer and fractional Doppler cases over linear time-varying (LTV) channels.
Moreover, we prove that IM enables AFDM to have stronger diversity protection when the full diversity condition is not satisfied.
Asymptotically tight upper bounds on the average bit error rates (BERs) of the proposed schemes with maximum-likelihood (ML) detection are derived in closed-form.
Furthermore, we propose a low-complexity double-layer message passing (DLMP) algorithm for practical large-dimensional signal detection in the proposed CDD-AFDM-IM systems.
Comparison with existing detections shows that the proposed DLMP algorithm achieves a better tradeoff between the BER performance and the computational complexity.
Finally, BER simulation results confirm that our proposed CDD-AFDM-IM schemes with both the ML and DLMP detections outperform the benchmark schemes over the LTV channels.
\end{abstract}

\section{Introduction}
\IEEEPARstart{N}{ext-generation} wireless communication systems (beyond 5G and 6G) will further extend the breadth and depth of communication networks coverage to realize the Internet of Everything (IoE)~\cite{6824752,9861699,9205980}.
One of the key challenges in this vision is the rapid relative movement between the transmitter and receiver, which results in a severe Doppler spreading effect in the wireless channel~\cite{9689960,7470933,7295466}.
The orthogonal frequency division multiplexing (OFDM) technique benefits from the use of cyclic prefix (CP) to avoid inter-symbol interference (ISI), which is an effective solution for time-dispersive channels and has been widely used in 4G and 5G systems~\cite{5635467}.
Nevertheless, the Doppler spreading effect leads to frequency dispersion, which destroys the orthogonality of OFDM subcarriers and degrades the transmission reliability~\cite{9849114}.
Therefore, it is imperative to develop new waveforms for next-generation communication systems to cope with the linear time-varying (LTV) channels.

So far, several new modulation waveforms have been developed for LTV channels.
As a representative, the orthogonal time-frequency space (OTFS) technique modulates the transmitted symbols over the delay-Doppler domain and then transforms the transmitted symbols to the time-frequency domain by employing the inverse symplectic finite Fourier transform (ISFFT)~\cite{7925924,10152009,10159363}.
It enables the transmission symbols to be multiplexed over the whole time-frequency domain, such that a potential full time-frequency diversity gain can be obtained.
A number of studies have shown that the OTFS technique can achieve an excellent performance gain compared to the OFDM technique over the LTV channel~\cite{8686339,10147252,10268002}.
Another approach to cope with LTV channels is to use chirp signals whose frequency varies with time~\cite{7050366,10146020}.
Thanks to this property, the chirp signals are very sensitive to frequency dispersion caused by Doppler shifts. 
Orthogonal chirp-division multiplexing (OCDM) is an example of this approach~\cite{7523229}. It is based on the discrete Fresnel transform, which shows better performance than OFDM over LTV channels. However, the diversity gain that can be obtained by the OCDM scheme is limited by particular channel delay-Doppler profiles.

Recently, based on the discrete affine Fourier transform (DAFT) that generalizes the discrete Fourier transform and discrete Fresnel transform, the affine frequency division multiplexing (AFDM) scheme has been proposed~\cite{9473655,9562168}.
By adjusting the parameters of the DAFT to accommodate both the maximum delay spread and maximum Doppler shift, the AFDM scheme can separate all channel paths in the {discrete affine Fourier (DAF)} domain, providing a complete delay-Doppler channel representation.
Therefore, the AFDM scheme can achieve a full diversity gain over the LTV channel, and significantly outperforms the traditional OFDM and OCDM schemes.
Moreover, although the performance of AFDM is similar to OTFS, it leads to a lower pilot overhead than OTFS.
This is due to the fact that AFDM gives rise to a one-dimensional representation for the delay-Doppler channel rather than a two-dimensional representation as in OTFS~\cite{10087310}.

Currently, AFDM research is still in its infancy.
There are few studies aiming at improving its performance or investigating its possible applications.
For example, a single-pilot and a multiple-pilot assisted channel estimation schemes were proposed for AFDM in~\cite{9880774}, where the advantage of the AFDM scheme in pilot consumption with respect to OTFS is confirmed.
An AFDM-based integrated sensing and communications (ISAC) system was studied in~\cite{9940346}, which shows that the AFDM-ISAC system can maintain excellent sensing performance even with a large Doppler shift.
For the detection of AFDM signals, by placing null symbols in the DAF domain to truncate the channel matrix into a sparse matrix, a low-complexity minimum-mean-square-error (MMSE) and a maximum ratio combining (MRC) receivers were proposed in~\cite{9746329}.
However, the placing of null symbols reduces the spectral efficiency (SE) of the AFDM~system.

Index modulation (IM) is a novel modulation technique that relies on the activation/inactivation of transmission resources to convey information without consuming additional energy~\cite{Wen2017IM,7509396,9761226}, which has been widely investigated in OFDM~\cite{Wen2021IM,6587554}, OTFS~\cite{10159363}, and multiple-input multiple-output (MIMO)~\cite{8765384} systems.
Attracted by the IM potential, more recently the authors in~\cite{tao2023affine} and~\cite{10342712} attempted to incorporate IM into the AFDM system.
The results revealed that the IM-assisted AFDM system does not always outperform the conventional AFDM system, which depends on the IM design and the channel setting. However, specific design guidelines for AFDM-IM with outstanding performance are missing. On the other hand, how to extend the AFDM-IM scheme to multiple-antenna scenarios with full transmit diversity is not clear nor has been investigated in the literature.

Against this background, this paper investigates a high-reliability IM scheme under the AFDM framework with multiple transmit antennas (TAs).
Our work is motivated by the fact that OFDM systems with the cyclic delay diversity (CDD) technique~\cite{10193343.1} have already demonstrated desired performance with the transmit diversity gain in static communication scenarios. We further extend the CDD technique to AFDM systems and analyze its performance in high mobility communication scenarios.
To the best of our knowledge, only~\cite{10193343} has given a simple cyclic delay-Doppler shift AFDM scheme, but it is not clear how the chirp parameters and the cyclic delay intervals of the adjacent TAs are chosen to achieve the full diversity. The lack of specific performance analysis and low-complexity efficient detectors in [33] limits the practical application of the CDD-AFDM scheme.
Different from the previous works~\cite{10193343.1,10193343}, we analyze the cyclic delay conditions in AFDM systems to achieve the full spatial diversity.
The specific requirements of the cyclic delay intervals for the adjacent TAs in both integer and fractional Doppler cases over the LTV channels are given.
Moreover, we carefully integrate the customized IM techniques with the multiple TAs AFDM framework to further enhance the transmission reliability.
We also develop a low-complexity detector for the proposed CDD-AFDM-IM schemes.
Specifically, the contributions of this paper can be summarized as follows.
\begin{itemize}
\item
We propose two novel CDD-assisted AFDM-IM schemes, termed as {\em CDD-AFDM-IM-I} and {\em CDD-AFDM-IM-II}, which improve the transmission reliability and energy efficiency of the AFDM system.
We discuss the full diversity conditions for the proposed CDD-AFDM-IM schemes.
Moreover, based on the optimal maximum likelihood (ML) detection, we derive closed-form bit error rate (BER) upper bounds on the proposed CDD-AFDM-IM schemes, through which the corresponding diversity orders are characterized.

\item
We develop a low-complexity double-layer message passing (DLMP) detection algorithm for the CDD-AFDM-IM schemes to facilitate their implementation in practice.
In the proposed algorithm, we design a two-layer message-passing structure to jointly detect the probability of modulation symbols and IM activation patterns.
The proposed DLMP algorithm is verified to be effective in both integer Doppler and fractional Doppler cases over LTV channels.

\item
Simulation results verify the tightness of the derived BER upper bounds in the high signal-to-noise ratio (SNR) regions. Our results also demonstrate that the proposed DLMP algorithm achieves a better tradeoff between BER performance and computational complexity than the benchmarks.
Moreover, we show that the CDD-AFM-IM schemes have a higher diversity order protection than the CDD-AFDM scheme when the full diversity condition is not satisfied. The above characteristics make the proposed CDD-AFDM-IM schemes particularly suitable for high-mobility communications with a large Doppler~shift.
\end{itemize}

The remainder of this paper is organized as follows.
Section~\ref{section:SysMod} presents the system model of the proposed CDD-AFDM-IM schemes.
Section~\ref{section:diversity} analyzes the parameter settings of the proposed scheme to obtain the full diversity.
Section~\ref{section:IM} illustrates the main ideas of the two proposed IM schemes with transmit diversity and analyzes their corresponding BER upper bounds based on ML detection.
The proposed low-complexity DLMP algorithm is described in Section~\ref{section:algorithm}.
In Section~\ref{section:Simulation}, simulation results are provided and discussed.
Finally, conclusions are drawn in Section~\ref{section:Conclusion}.

$Notations$: $(\cdot)^{T}$ and $(\cdot)^{H}$ represent the transpose and Hermitian transpose, respectively. $\left\lfloor \cdot  \right\rfloor $ represents the floor function.
$C(n,m)$ represents the binomial coefficient, which means that $m$ elements are randomly selected from $n$ elements.
$[\cdot]_{N}$ and $\left\| {\cdot} \right\|$ represent modulo $N$ and Euclidean norm, respectively.
$\rm diag(\cdot)$ transforms a vector into a diagonal matrix. $\mathbb{C}^{M \times N}$ and ${\bf{I}}_{N}$ are an $M\times N$ matrix with complex entries and an $N\times N$ identity matrix, respectively.
${\mathcal{CN}}(0,{\sigma}^2)$ represents the complex Gaussian distribution with zero mean and ${\sigma}^2$ variance.
${\rm rank}(\cdot)$ represents the rank of a matrix. $\lvert{\cdot\rvert}$ returns the absolute value. $\delta(\cdot)$ is the dirac delta function. $\otimes$ represents the convolution operator.
\begin{figure*}[t]
	\center
   \includegraphics[width=6.0in,height=2.5in]{{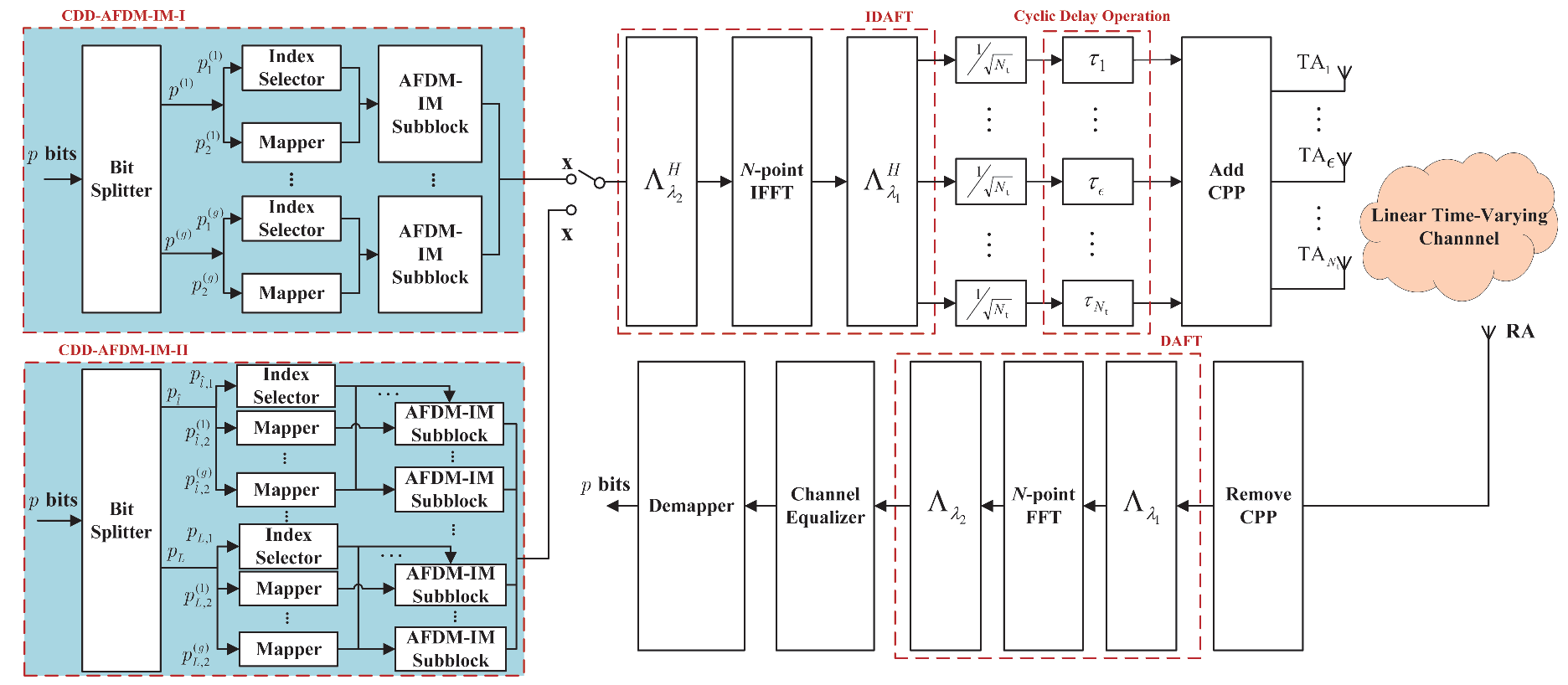}}
	\vspace{-0.2cm}
	\caption{{Block diagram of the transceiver structure for the proposed CDD-AFDM-IM schemes.}}
	\label{system-model}  
	\vspace{-0mm}
\end{figure*}
\section{System Model}
\label{section:SysMod}
In this section, we describe the system model of the proposed CDD-AFDM-IM schemes, where the corresponding transceiver structure is shown in Fig.~\ref{system-model}{\footnote{{This paper primarily focuses on a single-user scenario. In the future, we will consider the design of AFDM schemes in multi-user scenarios to provide a more comprehensive assessment of system performance.}}}.
The transmitter is equipped with ${N}_{\rm t}$ TAs.
The channel between the ${{\rm TA}_{\epsilon}}$ $(\epsilon=1,\ldots,{N}_{\rm t})$ and the receive antenna (RA) is modeled as an LTV channel.
We assume that the TAs are sufficiently spaced such that the fading of each TA is individually independent.
\vspace{-4mm}
\subsection{Transmitter}
{As shown in Fig.~\ref{system-model}, the transmitted signal generation at the transmitter will pass through four steps, i.e., IM mapping, IDAFT operation, cyclic delay operation, and adding chirp periodic prefix (CPP).
We first perform the IM and constellation symbol mapping in the DAF domain. The information bits are not only carried by the modulated symbols but also by the activation states of the chirp subcarriers in the DAF domain. Consider an AFDM symbol consisting of $N$ chirp subcarriers in the DAF domain. We define the transmitted symbol after our proposed IM mapping in the DAF domain as ${\bf{x}}\in{{\mathbb{C}}^{N\times1}}$.
Note that we proposed two IM mapping rules, and the purpose of this step is to generate the modulation symbol $\bf x$, which is independent of the subsequent operations. Therefore, the detailed description of the IM mapping rules is discussed in Section~\ref{section:IM}.}
Subsequently, applying the inverse (I-) DAFT, the transmitted signal in the time domain can be written as
\begin{equation}\label{AFDM.1}
{s}(t)=\frac{1}{\sqrt{N}}
\sum\limits_{u=0}^{N-1}{{{x}{[u]}}}{{e}^{j2\pi \left( {{\lambda }_{2}}{{u}^{2}}+\frac{u}{NT}t+\frac{{{\lambda }_{1}}}{{{T}^{2}}}{{t}^2} \right)}},
\end{equation}
where ${x}{[u]}$ is the $u$-$\rm{th}$ subsymbol in the DAF domain, $T$ is the sampling interval, and ${\lambda}_{1}$ and ${\lambda}_{2}$ are two adjustable parameters in IDAFT.
Let us define $t={\bar{u}}T$, ${\bar{u}}=0,\ldots,N-1$. Then,~\eqref{AFDM.1} is sampled to obtain the following matrix form
\begin{equation}\label{AFDM.2}
  {\bf s}={\bf\Lambda}_{{{\lambda}_{1}}}^{H}{{\bf F}^{H}}{\bf\Lambda}_{{{\lambda}_{2}}}^{H}{\bf x},
\end{equation}
where ${\bf\Lambda}_{{{\lambda_{1}}}}\!=\!{\rm diag}(e^{-j2\pi{\lambda_1}{{0}^2}}\!,\!\ldots\!,\!e^{-j2\pi {\lambda_1}{{\bar u}^2}}\!,\!\ldots\!,\!e^{-j2\pi {\lambda_1}{{(N-1)}^2}})$, ${\bf\Lambda}_{{{\lambda_{2}}}}\!=\!{\rm diag}(e^{-j2\pi{\lambda_2}{{0}^2}}\!,\!\ldots\!,\!e^{-j2\pi {\lambda_2}{{u}^2}}\!,\!\ldots\!,\!e^{-j2\pi {\lambda_2}{{(N-1)}^2}})$, ${{\bf F}\in{\mathbb{C}}^{N\times N}}$ is the $N$-point normalized discrete Fourier transform (DFT) matrix with ${\bf F}[\bar{u},{u}]=\frac{1}{\sqrt{N}}{e^{-j2\pi {\bar{u}}{u}/N}}$, and $u,{\bar u}\in\{0,1,\ldots,N-1\}$.

{To achieve transmit antenna diversity, we apply a cyclic delay operation on each TA. To ensure that the receiver can separate all paths, we need to carefully design the cyclic delay length of each antenna.
Therefore, the transmitted signal of the $\epsilon$-$\rm th$ TA is given by
\begin{equation}\label{3}
{s}(t\!-\!{\tau}_{\epsilon})\!=\!\!\frac{1}{\sqrt{N{N}_{\rm t}}}\!\!
\sum\limits_{u=0}^{N-1}\!\!{{{x}{[u]}}}{{e}^{j2\pi \left( {{\lambda }_{2}}{{u}^{2}}+\frac{u}{NT}(t-{\tau}_{\epsilon})+\frac{{{\lambda }_{1}}}{{{T}^{2}}}({{{t}-{\tau}_{\epsilon}})^2}\! \right)}},
\end{equation}
where ${\tau}_{\epsilon}$ represents the cyclic delay length of the $\epsilon$-$\rm th$ TA.
The parameters $\lambda_1$, $\lambda_2$ and ${\tau}_{\epsilon}$ can be adjusted according to the delay and Doppler distribution features over the LTV channel.
This approach enhances spatial diversity, ensuring that all channel paths are resolvable in the effective channel matrix, thereby enabling the proposed CDD-AFDM-IM schemes to achieve maximum full diversity.}
In Section~\ref{section:diversity}, we discuss the specific parameter settings of the proposed CDD-AFDM-IM schemes in details.
Unlike in OFDM, to avoid inter-block interference, the CPP~\cite{10087310} is added to the transmitted signal of each TA.
Without loss of generality, we assume that the length of CPPs are greater than the maximum channel delay spread. At last, the transmitted signal is sent to the receiver through the LTV channel.
\vspace{-4mm}
\subsection{Channel Model}
Consider an LTV channel with $P\ge1$ paths, and the channel impulse response (CIR) between the ${\epsilon}$-$\rm{th}$ TA and RA can be written as
\begin{equation}\label{AFDM.3}
h_{\epsilon}(\tau ,\nu )=\sum\limits_{\ell=1}^{P}{{{h}_{{\epsilon},\ell}}\delta (\tau -{{\tau }_{{\epsilon},\ell}})}{{e}^{j2\pi {{\nu }_{{\epsilon},\ell}}t}},
\end{equation}
where ${h}_{{\epsilon},\ell}\sim\mathcal{CN}(0,1/P)$, ${\tau}_{{\epsilon},\ell}\in[0,{\tau}_{\rm max}]$, and ${\nu}_{{\epsilon},\ell}\in[-{\nu}_{\rm max},{\nu}_{\rm max}]$ represent the CIR coefficient, actual delay spread, and actual Doppler spread index associated with the $\ell$-$\rm th$ path for the ${\epsilon}$-$\rm{th}$ TA, respectively.
The normalized delay shift and Doppler shift of the $\ell$-$\rm th$ path are given by ${l}_{{\epsilon},\ell}={\tau}_{{\epsilon},\ell}{\Delta f}\in[0,{l}_{\rm{max}}]$ and ${\alpha_{{\epsilon},\ell}}={\nu}_{{\epsilon},\ell}{NT}$, where ${\Delta f}$ is the chirp subcarrier spacing and $T{\Delta f}=1$.
We define ${\alpha }_{{\epsilon},\ell}={\alpha }_{{\rm int},{\epsilon},\ell}+{\alpha }_{{\rm fra},{\epsilon},\ell}$ with ${\alpha }_{{\rm int},{\epsilon},\ell}\in[-{\alpha }_{\rm{max}},{\alpha }_{\rm{max}}]$ and $-\frac{1}{2}<{\alpha }_{{\rm fra},{\epsilon},\ell}<\frac{1}{2}$, which refer to the integer normalized Doppler shift and fractional normalized Doppler shift, respectively.
It is worth mentioning that this channel model is a generalized model that allows each delay tap to have different Doppler shift values, i.e., ${\ell}_{1},{\ell}_{2}=1,\ldots,P$, ${\tau}_{{\epsilon},{\ell}_{1}}={\tau}_{{\epsilon},{\ell}_{2}}$, and ${\alpha}_{{\epsilon},{\ell}_{1}}\ne{\alpha}_{{\epsilon},{\ell}_{2}}$~\cite{9473655,9562168,10087310}.
Therefore, for integer Doppler shifts, we can calculate the maximum possible number of paths for each antenna as ${P}_{\rm max}=({{l}_{\rm max}+1})(2{\alpha}_{\rm max}+1)$.
\vspace{-4mm}
\subsection{Receiver}
At the receiver, after removing the CPP, the received signal in the time domain can be written as
\begin{equation}\label{5}
r(t)=\sum\limits_{\epsilon=1}^{{N}_{\rm t}}\sum\limits_{\ell=1}^{P}{{{h}_{\epsilon,\ell}}s(t-{{\tau }_{\epsilon,\ell}}-{{\tau }_{\epsilon}}){{e}^{j2\pi {{\nu }_{\epsilon,\ell}}t}}}+w(t),
\end{equation}
where $w(t)\sim\mathcal{CN}(0,N_0)$ is the additive and complex Gaussian random noise. Then, the received time-domain signal is transformed back to the DAF domain by using DAFT, and one can obtain the $\bar v$-$\rm th$ received symbol in the DAF domain, which is given by~\eqref{6.1}, as shown at the top of this page.
\begin{figure*}[ht]
\begin{align}\label{6.1}
   {{{y}}{{{[\bar v]}}}}\!=&\!\frac{1}{\sqrt{N}T}\!\!\int_{0}^{NT}\!\!{r(t){{e}^{-j2\pi \left( {{\lambda }_{2}}{{{\bar{v}}}^{2}}+\frac{{\bar{v}}}{NT}t+\frac{{{\lambda }_{1}}}{{{T}^{2}}}{{t}^{2}} \right)}}}dt \nonumber\\
  =&\!\frac{1}{NT}\!\!\sum\limits_{\epsilon=1}^{{N}_{\rm t}}\!\sum\limits_{\ell=1}^{P}\!\frac{{{h}_{\epsilon,\ell}}}{\sqrt{{N}_{\rm t}}}\!\!\sum\limits_{v=0}^{N-1}\!\!{{{x}{[v]}}{{e}^{j2\pi \left[ {{\lambda }_{2}}({{v}^{2}}-{{{\bar{v}}}^{2}})-\frac{v\left({{\tau }_{\epsilon,\ell}}+{{\tau }_{\epsilon}}\right)}{NT}+\frac{{{\lambda }_{1}}\left({{\tau }_{\epsilon,\ell}}+{{\tau }_{\epsilon}}\right)^{2}}{{{T}^{2}}}
   \right]}}}\!\!\!
  \int_{0}^{NT}\!\!\!\!{{{e}^{j\frac{2\pi }{NT}\left[v-\left( \bar{v}-NT{{\nu }_{\epsilon,\ell}}+2\frac{{{\lambda }_{1}}}{T}N\left({{\tau }_{\epsilon,\ell}}+{{\tau }_{\epsilon}}\right) \right) \right]t}}}dt+{{{\overline{w}}}{[\bar v]}}
\end{align}
\hrulefill
\vspace{-4mm}
\end{figure*}
Let us define ${{\tau }_{\epsilon}}=T{{l}_{\epsilon}}$, and~\eqref{6.1} can be rewritten as
\begin{align}\label{7}
   {{{{y}}}{{[{\bar v}]}}}=&\frac{1}{N}\sum\limits_{\epsilon=1}^{{N}_{\rm t}}\sum\limits_{\ell=1}^{P}\frac{{{h}_{\epsilon,\ell}}}{\sqrt{{N}_{\rm t}}}\sum\limits_{v=0}^{N-1}{{{{x}}{[v]}}{{e}^{j\frac{2\pi}{N}\zeta}}}
 {\mathcal{F}}_{\epsilon,\ell}[\bar{v},{v}]
 +{{{\overline{w}}}{[\bar v]}},
\end{align}
where ${{{\overline{w}}}{[\bar v]}}$ is the filtered output of ${{w}(t)}$ in the DAF domain, $\zeta=N{{\lambda }_{2}}({{v}^{2}}-{{{\bar{v}}}^{2}})-v{({{l}_{\epsilon,\ell}}+{{l}_{\epsilon}})}+N{{\lambda }_{1}}({{l}_{\epsilon,\ell}}+{{l}_{\epsilon}})^{2}$, and
\begin{align}\label{7.1}
   {{\mathcal{F}}_{\epsilon ,\ell }}[\bar{v},v]&=\sum\limits_{n=0}^{N-1}{{{e}^{j\frac{2n\pi }{N}\left[ v-\left( \bar{v}-{{\alpha }_{\epsilon ,\ell }}+2N{{\lambda }_{1}}\left( {{l}_{\epsilon ,\ell }}+{{l}_{\epsilon }} \right) \right) \right]}}} \nonumber\\
 & =\frac{{{e}^{j2\pi \left[ v-\left( \bar{v}-{{\alpha }_{\epsilon ,\ell }}+2N{{\lambda }_{1}}\left( {{l}_{\epsilon ,\ell }}+{{l}_{\epsilon }} \right) \right) \right]}}-1}{{{e}^{j\frac{2\pi }{N}\left[ v-\left( \bar{v}-{{\alpha }_{\epsilon ,\ell }}+2N{{\lambda }_{1}}\left( {{l}_{\epsilon ,\ell }}+{{l}_{\epsilon }} \right) \right) \right]}}-1}.
\end{align}
The DAF domain input-output relation in~\eqref{7} can be vectorized as
\begin{equation}\label{8}
{\bf y}={\bf H}_{\rm eff}{\bf x}+{\overline{\bf w}},
\end{equation}
where ${\bf x}$, ${\bf y}\in{{\mathbb C}^{N\times 1}}$, $\overline{\bf w}\in{{\mathbb C}^{N\times 1}}$ is the noise vector, ${\bf H}_{\rm eff}=\frac{1}{\sqrt{{N}_{\rm t}}}\sum\nolimits_{\epsilon=1}^{{N}_{\rm t}}\sum\nolimits_{\ell=1}^{P}{h}_{\epsilon,\ell}{\bf H}_{\epsilon,\ell}$ represents the effective channel matrix, and
\begin{equation}\label{8.1}
{\bf H}_{\epsilon,\ell}=\frac{1}{N}{{e}^{j\frac{2\pi}{N}\zeta}}{\mathcal{F}}_{\epsilon,\ell}[\bar{v},{v}].
\end{equation}
Finally, after the channel equalizer and demapper, we can obtain the estimation of the transmitted information bits.
\section{Parameter Settings and Analysis}
\label{section:diversity}
As can be found from~\eqref{7.1}, we can flexibly adjust the parameters $\lambda_1$, $\lambda_2$ and ${l}_{\epsilon}$ according to the delay and doppler distribution features and thus realize the reconfiguration of the channel to ensure the optimal performance. Therefore, in this section we will discuss the settings of these parameters in detail.
There is a difference in the output of \eqref{8.1} when considering the integer Doppler and fractional Doppler cases.
This will lead to some differences in the parameter settings of the proposed CDD-AFDM-IM schemes for full diversity gain achievement.
Therefore, we discuss corresponding parameter settings separately.
First, for the integer Doppler shifts, the expression for ${\bf H}_{\epsilon,\ell}$ is given by
\begin{equation}\label{9}
  {{H}_{\epsilon,\ell}}[\bar{v},{v}]\!=\!\left\{ \begin{array}{*{35}{l}}
   \!{{e}^{j\frac{2\pi }{N}\zeta}}, \!&v={{\left[ \bar{v}+{\rm loc}_{\epsilon,\ell} \right]}_{N}},  \\
   \!0, \!& {\rm otherwise},
\end{array} \right.
\end{equation}
where ${\rm loc}_{\epsilon,\ell}=-{{\alpha }_{\epsilon,\ell}}+2N{{\lambda }_{1}}({{l}_{\epsilon,\ell}}+{{l}_{\epsilon}})$, ${\rm loc}_{\epsilon,\ell}\in{{\mathbb A}_{\epsilon,\ell}}$, and ${\mathbb A}_{\epsilon,\ell}=\{{-{{\alpha }_{\rm max}}+2N{{\lambda }_{1}}({{l}_{\epsilon,\ell}}+{{l}_{\epsilon}}),\ldots,{{\alpha }_{\rm max}}+2N{{\lambda }_{1}}({{l}_{\epsilon,\ell}}+{{l}_{\epsilon}})}\}$.~We can find that each row of ${{\bf H}_{\epsilon,\ell}}$ has ${2{\alpha}_{\rm max}+1}$ non-zero elements.

For fractional Doppler shifts, the interference caused by each fractional Doppler basically extends over the whole DAF domain with a position as the center, and the interference decreases as one moves away from the center point.
Therefore, in the matrix ${\bf H}_{\epsilon,\ell}$, we consider that the main Doppler interference exists within an interval of size $2{k}_{\alpha}+1$ centered on $v={{\left[ \bar{v}+{\rm loc}_{\epsilon,\ell} \right]}_{N}}$~\cite{10087310},
where ${k}_{\alpha}<|v-{{\left[ \bar{v}+{\rm loc}_{\epsilon,\ell} \right]}_{N}}|$ is a nonnegative integer. Therefore, the expression of the matrix ${\bf H}_{\epsilon,\ell}$ can be written as
\begin{equation}\label{9.1}
  {{H}_{\epsilon,\ell}}[\bar{v},{v}]\!=\!\left\{ \begin{array}{*{35}{l}}
   \!\frac{1}{N}{{e}^{j\frac{2\pi}{N}\zeta}}{\mathcal{F}}_{\epsilon,\ell}[\bar{v},{v}], \!& v={{\left[ \bar{v}+{\rm loc}_{{\rm fra},\epsilon,\ell} \right]}_{N}}, \\
   \!0, \!& {\rm otherwise},
\end{array} \right.
\end{equation}
where ${\rm loc}_{\epsilon,\ell} -{k}_{\alpha}\le{\rm loc}_{{\rm fra},\epsilon,\ell}\le{\rm loc}_{\epsilon,\ell} +{k}_{\alpha}$.
Since ${\rm loc}_{\epsilon,\ell}\in{{\mathbb A}_{\epsilon,\ell}}$, we can define ${\rm loc}_{{\rm fra},\epsilon,\ell}\in{{\mathbb B}_{\epsilon,\ell}}$ and ${\mathbb B}_{\epsilon,\ell}=\{\!{-\!{{\alpha }_{\rm max}}\!\!-\!\!{k}_{\alpha}\!+\!2N{{\lambda }_{1}}\!({{l}_{\epsilon,\ell}}\!+\!{{l}_{\epsilon}})\!,\ldots,\!{{\alpha }_{\rm max}}\!+\!{k}_{\alpha}\!+\!2N{{\lambda }_{1}}\!({{l}_{\epsilon,\ell}}\!+\!{{l}_{\epsilon}})}\}$.

To ensure that our CDD-AFDM-IM systems achieve full diversity, it is necessary to ensure that the non-zero elements of ${{\bf H}_{\epsilon,\ell}}$ and ${{\bf H}_{{\epsilon}^{'},\ell^{'}}}$ do not overlap, where ${\epsilon}\ne{\epsilon}^{'}\in[1,{N_t}]$ and ${\ell}\ne{\ell}^{'}\in[1,{P}]$.

{\bf{\em DAFT parameter settings:}}
First, we determine the parameter settings for a single antenna to obtain a full diversity gain.
Specifically, for the ${\epsilon}$-$\rm th$ TA,
we need to ensure that the intersection of ${\mathbb A}_{\epsilon,\ell}$ and ${\mathbb A}_{\epsilon,\ell^{'}}$ or ${\mathbb B}_{\epsilon,\ell}$ and ${\mathbb B}_{\epsilon,\ell^{'}}$ is empty, i.e.,
\begin{equation}\label{10}
{{\mathbb A}_{\epsilon,\ell}}\cap{{\mathbb A}_{\epsilon,\ell^{'}}}=\varnothing,
\end{equation}
or
\begin{equation}\label{10.1}
{{\mathbb B}_{\epsilon,\ell}}\cap{{\mathbb B}_{\epsilon,\ell^{'}}}=\varnothing.
\end{equation}
We assume $\ell<\ell^{'}$, and the constraint in~\eqref{10} can be converted to
\begin{align}\label{11}
  & {{\alpha }_{\rm{max}}}+2N{{\lambda }_{1}}({{l}_{\epsilon ,\ell }}+{{l}_{\epsilon }})<-{{\alpha }_{\rm{max}}}+2N{{\lambda }_{1}}({{l}_{\epsilon ,{{\ell }^{'}}}}+{{l}_{\epsilon }}) \nonumber\\
  \Rightarrow &\frac{2{{\alpha }_{\rm{max}}}}{{{l}_{\epsilon ,\ell^{'} }}-{{l}_{\epsilon ,{{\ell }}}}}<2N{{\lambda }_{1}}.
\end{align}
Similarly, we can convert the constraint in~\eqref{10.1} as
\begin{align}\label{11.1}
  & {{\alpha }_{\rm{max}}}\!+\!{k}_{\alpha}\!+\!2N{{\lambda }_{1}}({{l}_{\epsilon ,\ell }}\!+\!{{l}_{\epsilon }})\!<\!\!-\!{{\alpha }_{\rm{max}}}\!-\!{k}_{\alpha}\!+\!2N{{\lambda }_{1}}({{l}_{\epsilon ,{{\ell }^{'}}}}\!+\!{{l}_{\epsilon }})\nonumber\\
  \Rightarrow &\frac{2({{\alpha }_{\rm{max}}}\!+\!{k}_{\alpha})}{{{l}_{\epsilon ,\ell^{'} }}\!-\!{{l}_{\epsilon ,{{\ell }}}}}<2N{{\lambda }_{1}}.
\end{align}
Therefore, we set ${\lambda }_{1}=({2{{\alpha }_{\rm{max}}}+1})/{2N}$ in the integer Doppler case and ${\lambda }_{1}=({2{{\alpha }_{\rm{max}}}+2{k}_{\alpha}+1})/{2N}$ in the fractional Doppler case.
We also set ${\lambda}_{2}$ as an arbitral irrational number or a rational number sufficiently smaller than $1/2N$~\cite{10087310} to satisfy the full diversity condition for a single~TA.

{\bf{\em Cyclic delay interval setting:}}
Then, for the case of multiple TAs, we also require that the non-zero elements of the effective channel matrices do not overlap.
For the integer Doppler shifts, we denote ${\rm loc}_{\epsilon}\in{{\mathbb A}_{\epsilon}}$, and ${\mathbb A}_{\epsilon}=\{{-{{\alpha }_{\rm max}}+2N{{\lambda }_{1}}{{l}_{\epsilon}},\ldots,{{\alpha }_{\rm max}}+2N{{\lambda }_{1}}({{l}_{\rm max}}+{{l}_{\epsilon}})}\}$.
For the fractional Doppler shifts, we denote ${\rm loc}_{{\rm frac},\epsilon}\in{{\mathbb B}_{\epsilon}}$, and ${\mathbb B}_{\epsilon}=\{{-{{\alpha }_{\rm max}}\!-\!{k}_{\alpha}\!+\!2N{{\lambda }_{1}}{{l}_{\epsilon}},\ldots,{{\alpha }_{\rm max}}\!+\!{k}_{\alpha}\!+\!2N{{\lambda }_{1}}({{l}_{\rm max}}+{{l}_{\epsilon}})}\}$.
To enable our CDD-AFDM-IM schemes to further achieve the full diversity of multiple transmit antennas, the intersection of sets ${{\mathbb A}_{\epsilon}}$ and ${{\mathbb A}_{{\epsilon}^{'}}}$ or ${{\mathbb B}_{\epsilon}}$ and ${{\mathbb B}_{{\epsilon}^{'}}}$ must be empty, i.e., ${{\mathbb A}_{\epsilon}}\cap{{\mathbb A}_{{\epsilon}^{'}}}=\varnothing$ or ${{\mathbb B}_{\epsilon}}\cap{{\mathbb B}_{{\epsilon}^{'}}}=\varnothing$.
Similarly, assuming $l_{\epsilon}<l_{\epsilon^{'}}$, we can obtain the conditions of achieving full transmit diversity, respectively, for the integer and fractional Doppler cases as follows
\begin{align}\label{13}
  \frac{2{{\alpha }_{\rm{max}}}}{2N{\lambda }_{1}}+{l}_{\rm max}<{{l}_{{\epsilon}^{'}}}-{{l}_{{\epsilon}}},
\end{align}
and
\begin{align}\label{13.1}
  \frac{2({{\alpha }_{\rm{max}}}+{k}_{\alpha})}{2N{\lambda }_{1}}+{l}_{\rm max}<{{l}_{{\epsilon}^{'}}}-{{l}_{{\epsilon}}}.
\end{align}
Thus, we can obtain the minimum normalized cyclic delay interval between adjacent TAs of the proposed CDD-AFDM-IM schemes as ${\Delta}_{\rm min}={{l}_{{\epsilon}^{'}}}-{{l}_{{\epsilon}}}$, where the corresponding conditions required for integer and fractional Doppler cases are respectively expressed as follows
\begin{align}\label{14}
  {\Delta}_{\rm min}>{l}_{\rm max}+\frac{2{{\alpha }_{\rm{max}}}}{2N{\lambda }_{1}},
\end{align}
and
\begin{align}\label{14.1}
  {\Delta}_{\rm min}>{l}_{\rm max}+\frac{2({{\alpha }_{\rm{max}}}+{k}_{\alpha})}{2N{\lambda }_{1}}.
\end{align}
After determining the value of ${\lambda }_{1}$, we can set ${\Delta}_{\rm min}={l}_{\rm max}+1$ to realize the full diversity conditions for both the integer and fractional Doppler cases.

{\bf{\em Transmitted signal dimension setting:}}
In addition, the maximum number of possible paths for the proposed CDD-AFDM-IM schemes cannot be greater than the number of chirp symbols $N$ to ensure that the effective channel matrix ${\bf{H}}_{\rm{eff}}$ has enough space to accommodate all the channel delay, Doppler and antenna space paths.
Specifically, the full diversity conditions required for integer and fractional Doppler cases can be respectively expressed as
\begin{equation}\label{CDD-AFDM.1}
{({l_{\rm{max}}}+1)(2{{\alpha}_{\rm{max}}}+1){N}_{\rm{t}}}\le N,
\end{equation}
and
\begin{align}\label{14.3}
  {({l_{\rm{max}}}+1)(2{{\alpha}_{\rm{max}}}+2{{k}_{\alpha}+1})}{N_{\rm{t}}}\le N.
\end{align}
\begin{table}[t]
\caption{{An Example of the Proposed Schemes With $n=4$ and $m=2$. }}\label{table1}
\centering
\renewcommand\arraystretch{1.4}
\setlength{\tabcolsep}{6mm}{
\begin{tabular}{|c|c|}
\hline
Index Bits & ${\bf x}^{(i)}$ \\ \hline \hline
{[0~0]}  & {$[q^{(i)}_{1},~q^{(i)}_{2},~0,~0]^T$}           \\ \hline
{[0~1]}  & {$[0,~q^{(i)}_{1},~q^{(i)}_{2},~0]^T$}           \\ \hline
{[1~0]}  & {$[0,~0,~q^{(i)}_{1},~q^{(i)}_{2}]^T$}           \\ \hline
{[1~1]}  & {$[q^{(i)}_{1},~0,~0,~q^{(i)}_{2}]^T$}           \\ \hline
\end{tabular}}
\vspace{-2mm}
\end{table}
\section{Proposed IM-assisted AFDM Scheme With Transmit Diversity}
\label{section:IM}
In this section, we present the proposed IM-assisted AFDM scheme with transmit diversity. We first introduce the CDD-AFDM-IM-I scheme and then design the CDD-AFDM-IM-II scheme from the view of diversity analysis.
\subsection{Proposed CDD-AFDM-IM-I Scheme}
\label{section:IM.A}
To perform IM, we first divide the $N$ DAF domain subsymbols into $g$ groups, and the number of subsymbols in each group is $n=N/g$. Take the $i$-${\rm th}$ ($i=1,\ldots,g$) group as an example. Each group needs to carry ${p}^{(i)}=p/g$ bits, where $p$ denotes the total number of bits carried by the transmitted symbols, and the ${p}^{(i)}$ bits are further divided into ${p}^{(i)}_{1}$ index bits and ${p}^{(i)}_{2}$ modulated bits.
The index selector randomly selects $m$ out of $n$ chirp subcarriers to be active based on the input ${p}^{(i)}_{1}$ index bits, and the remaining $n-m$ chirp subcarriers are inactive. The $m$ active chirp subcarriers are used to carry the constellation symbols ${\bf{q}}^{(i)}=\{{q}^{(i)}_{1};\ldots;{q}^{(i)}_{m}\}$.
For example, when $n=4$ and $m=2$, the symbol vector embedded in the $i$-$\rm th$ group ${\bf{x}}^{(i)}\in{{\mathbb{C}}^{n\times1}}$ is given in Table~\ref{table1}, which has four possible choices.
Therefore, the number of index bits ${p}^{(i)}_{1}={\rm log}_2\left\lfloor C(n,m) \right\rfloor$, and the number of modulated bits ${p}^{(i)}_{2}=m{\rm log}_2(M)$, where $M$ is the cardinality of the signal constellation.
After obtaining the transmitted symbol $\bf{x}$ in the DAF domain, the subsequent operations are performed, as described in Section~\ref{section:SysMod}.

{\bf{\em Diversity order analysis:}}
For the proposed CDD-AFDM-IM-I scheme, the received signal in the DAF domain can be written as
\begin{equation}\label{CDD-AFDM.2}
 {\bf y}=\frac{1}{\sqrt{{N}_{\rm t}}}\sum\limits_{\epsilon=1}^{{N}_{\rm t}}\sum\limits_{\ell=1}^{P}{{{h}_{\epsilon,\ell}}{{\bf H}_{\epsilon,\ell}}{\bf x}+\overline{\bf w}={\bf\Upsilon}_{\bf x}}{\bf h}+\overline{\bf w},
\end{equation}
where ${\bf h}=\frac{1}{\sqrt{{N}_{\rm t}}}[{h}_{1,1},{h}_{1,2},\ldots,{h}_{{{N}_{\rm t}},P}]^{T}\in{\mathbb{C}}^{P{{N}_{\rm t}}\times1}$ is a path coefficient vector,
and ${\bf\Upsilon}_{\bf x}=[{{\bf H}_{1,1}{\bf x}},{{\bf H}_{1,2}{\bf x}},\ldots,{{\bf H}_{{{N}_{\rm t}},P}{\bf x}}]\in{{\mathbb{C}}^{N\times P{{N}_{\rm t}}}}$.
For the received signal in~\eqref{CDD-AFDM.2}, we can make a joint decision on the two information carrying units based on the ML detector, which can be formulated as
\begin{equation}\label{CDD-AFDM.3}
\left(\widetilde{\mathbf{k}},\widetilde{\mathbf{q}} \right)= \underset{\mathbf{k},\mathbf{q}}{\rm{arg~min }}\,{{\left\| {\bf y}-{{\bf\Upsilon}_{\bf x}}{\bf h} \right\|}^{2}},
\end{equation}
where ${\bf{k}}$ and ${\bf{q}}$ are the indices of the active chirp subcarriers and the $M$-ary modulation symbol, respectively.
For the $i$-$\rm{th}$ group, we have ${\bf k}^{(i)}=\{{k}^{(i)}_{1};\ldots;{k}^{(i)}_{m}\}$ and ${\bf q}^{(i)}=\{{q}^{(i)}_{1};\ldots;{q}^{(i)}_{m}\}$, where ${k}^{(i)}_{j}\in\{1,\ldots,n\}$, $j\in\{1,\ldots,m\}$, and ${q}^{(i)}_{j}$ is an $M$-PSK/QAM symbol.

{\em Lemma~1:} Define ${\bf{x}}_{i}$ and ${\bf{x}}_{j}$ as two different realizations of DAF domain transmitted vectors, and the difference matrix ${\bf\Upsilon}_{{\bf{x}}_{i}-{\bf{x}}_{j}}={\bf\Upsilon}_{{\bf{x}}_{i}}-{\bf\Upsilon}_{{\bf{x}}_{j}}$.
Then, the diversity order of the proposed scheme is characterized as the rank of the difference matrix ${\bf\Upsilon}_{{\bf{x}}_{i}-{\bf{x}}_{j}}$, as follows

\begin{equation}\label{CDD-AFDM.4}
d=\underset{{\bf{x}}_{i}, {\bf{x}}_{j}, {i\ne j}}{\rm{min}} {\rm rank}\left({\bf\Upsilon}_{{\bf{x}}_{i}-{\bf{x}}_{j}}\right).
\end{equation}
\begin{IEEEproof}
The diversity order is related to the pairwise error probability (PEP) in the high SNR region.
We assume that the perfect {channel state information (CSI)} is known at the receiver,
and thus, the conditional PEP, i.e., the probability of transmitting ${\bf x}_{i}$ but erroneously deciding on ${\bf x}_{j}$ is given by
\begin{align}\label{17}
  & {\rm Pr}\left({{\bf x}_{i}}\to {{\bf x}_{j}}|{\bf h} \right)
 \!=\!Q\left( \sqrt{\frac{{{\left\| \left( {{\bf\Upsilon }_{{{\bf x}_{i}}}}-{{\bf\Upsilon }_{{{\bf x}_{j}}}} \right){\bf h} \right\|}^{2}}}{{2{{N}_{0}}}} }\right),
\end{align}
where $Q\{\cdot\}$ is the tail distribution function of the standard Gaussian distribution.
According to~\cite{10159363}, we can approximate
\begin{equation}\label{18}
Q(x)\cong \frac{1}{12}{{e}^{-{{x}^{2}}/2}}+\frac{1}{4}{{e}^{-2{{x}^{2}}/3}}.
\end{equation}
Then, the unconditional PEP can be approximated by
\begin{equation}\label{19}
{\rm Pr}\left( {{\bf x}_{i}}\to {{\bf x}_{j}} \right)\cong {{E}_{{{\bf h}}}}\left\{ \frac{1}{12}{{e}^{-{{q}_{1}}\Phi }}+\frac{1}{4}{{e}^{-{{q}_{2}}\Phi }} \right\},
\end{equation}
where $q_1\!\!=\!\!1/4{N_0}$, $q_2\!\!=\!\!1/3{N_0}$,~and $\Phi={{{\left\| {\bf h}\left( {{\bf\Upsilon }_{{{\bf x}_{i}}}}-{{\bf\Upsilon }_{{{\bf x}_{j}}}} \right) \right\|}^{2}}}={{{\bf h}}}{\bf{A}}{{{\bf h}^{H}}}$.
Here, the matrix ${\bf A}={\bf{U}}{\bf{\Lambda}}{{\bf{U}}^{H}}$ is a Hermitian matrix, where ${\bf{U}}$ is unitary and ${\bf{\Lambda}}={\rm diag}\{\kappa _{1}^{2},\ldots,\kappa _{P{N}_{\rm{t}}}^{2}\}$ with $\kappa _{i}$ being the $i$-$\rm th$ singular value of the matrix ${\bf\Upsilon}_{{\bf{x}}_{i}-{\bf{x}}_{j}}={\bf\Upsilon}_{{\bf{x}}_{i}}-{\bf\Upsilon}_{{\bf{x}}_{j}}$.
Therefore, the unconditional PEP in~\eqref{19} can be calculated as
\begin{align}\label{21}
{\rm Pr}\left( {{\bf x}_{i}}\to {{\bf x}_{j}} \right)& \cong
\frac{1}{12}\prod\limits_{i=1}^{d}{\frac{1}{1+\frac{{{q}_{1}}\kappa _{i}^2}{{{{N}_{\rm t}}P}}}}+\frac{1}{4}\prod\limits_{i=1}^{d}{\frac{1}{1+\frac{{{q}_{2}}\kappa _{i}^2}{{{N}_{\rm t}}P}}},
\end{align}
where $d$ is the rank of the matrix ${\bf\Upsilon}_{{\bf{x}}_{i}-{\bf{x}}_{j}}$.
In the high SNR region,~\eqref{21} can be approximated as
\begin{equation}\label{22}
{\rm Pr}\left( {{\mathbf{x}}_{i}}\to {{\mathbf{x}}_{j}} \right)\approx \left({\prod\limits_{i=1}^{d}}\kappa _{i}^2\right)^{-1}\left(\frac{4^d}{12}+\frac{3^d}{4}\right){\rm SNR}^{-d},
\end{equation}
where SNR is defined as $1/{{N}_{0}}$. As a result, one can find that the diversity order of the proposed CDD-AFDM-IM-I scheme is the rank of the difference matrix ${\bf\Upsilon}_{{\bf{x}}_{i}-{\bf{x}}_{j}}$, which could be as high as ${\rm min}(N, P{N}_{\rm t})$.
\end{IEEEproof}
Moreover, based on the obtained PEP, we can calculate the average bit error probability (ABEP) upper bound on the proposed CDD-AFDM-IM-I scheme as
\begin{equation}\label{23}
{{\rm Pr}_{\rm ABEP}}\le\frac{1}{{{2}^{p}}p}\sum\limits_{{{\bf x}_{i}}}{\sum\limits_{{{\bf x}_{j}}}{{\rm Pr}\left( {{\bf x}_{i}}-{{\bf x}_{j}} \right)N\left( {{\bf x}_{i}},{{\bf x}_{j}} \right)}},
\end{equation}
where $N\left( {{\bf x}_{i}},{{\bf x}_{j}} \right)$ denotes the number of error bits for ${{\bf x}_{i}}$ when it is estimated as ${{\bf x}_{j}}$.

{\bf{\em Guidelines for IM design:}}
However, in the multiple TAs scenarios, the full diversity condition in~\eqref{CDD-AFDM.1} or~\eqref{14.3}
is difficult to satisfy for a large number of TAs as $N$ cannot be a large value due to the low-latency communication requirement, which results in diversity loss of modulated bits.
Fortunately, as will be shown later, IM can obtain an additional diversity gain which can compensate for the performance loss of multiple antenna AFDM system when the full diversity condition in~\eqref{CDD-AFDM.1} or~\eqref{14.3} is not satisfied.
Let us take an example to briefly illustrate the diversity protection feature of IM. Suppose $P=1$, $N_{\rm{t}}=2$, $N=4$, $g=1$, $m=1$, ${l}_{\rm{max}}=0$, and ${\alpha}_{\rm{max}}=1$, such that the full diversity condition in~\eqref{CDD-AFDM.1} is not satisfied.{\footnote{Here, we take the integer Doppler case as an example to clearly illustrate the design guidelines and advantages of IM, which also apply to the fractional Doppler case.}} We can express the difference matrix ${\bf\Upsilon}_{{\bf{x}}_{i}-{\bf{x}}_{j}}$ as
\begin{equation}\label{CDD-AFDM.5}
{\bf\Upsilon}_{{\bf{x}}_{i}-{\bf{x}}_{j}}=\left[{\bf{H}}_{1,1}\left({{\bf{x}}_{i}}-{{\bf{x}}_{j}}\right),{\bf{H}}_{2,1}\left({{\bf{x}}_{i}}-{{\bf{x}}_{j}}\right)\right].
\end{equation}
Based on the setting of delay and Doppler parameters, we set the normalized cyclic delay interval of the  adjacent TAs as $\Delta=\Delta_{\rm{min}}=1$.
According to~\eqref{9}, the forms of ${\bf{H}}_{1,1}$ and ${\bf{H}}_{2,1}$ are given by
\begin{equation}\label{CDD-AFDM.6}
{\bf{H}}_{1,1}=\left[ \begin{array}{*{35}{l}}
   1 & 0 & 0 & 0  \\
   0 & 1 & 0 & 0  \\
   0 & 0 & 1 & 0  \\
   0 & 0 & 0 & 1  \\
\end{array} \right],
\end{equation}
and
\begin{equation}\label{CDD-AFDM.7}
{\bf{H}}_{2,1}=\left[ \begin{array}{*{35}{l}}
   e^{j{\frac{\pi}{4}}} & 0 & 0 & 0  \\
   0 & e^{-j{\frac{\pi}{4}}} & 0 & 0  \\
   0 & 0 & e^{-j{\frac{3\pi}{4}}} & 0  \\
   0 & 0 & 0 & e^{-j{\frac{5\pi}{4}}}  \\
\end{array} \right].
\end{equation}
Then, based on~\eqref{CDD-AFDM.6} and~\eqref{CDD-AFDM.7}, we can rewrite~\eqref{CDD-AFDM.5} as
\begin{equation}\label{CDD-AFDM.8}
{\bf\Upsilon}_{{\bf{x}}_{i}}-{\bf\Upsilon}_{{\bf{x}}_{j}}=\left[ \begin{array}{*{35}{l}}
   {x}_{i,1}-{x}_{j,1} & \left({x}_{i,1}-{x}_{j,1}\right)e^{j{\frac{\pi}{4}}}   \\
   {x}_{i,2}-{x}_{j,2}, & \left({x}_{i,2}-{x}_{j,2}\right)e^{-j{\frac{\pi}{4}}}   \\
   {x}_{i,3}-{x}_{j,3} & \left({x}_{i,3}-{x}_{j,3}\right)e^{-j{\frac{3\pi}{4}}}  \\
   {x}_{i,4}-{x}_{j,4} & \left({x}_{i,4}-{x}_{j,4}\right)e^{-j{\frac{5\pi}{4}}}   \\
\end{array} \right].
\end{equation}
For the conventional CDD-AFDM system, the information bits are carried only by the modulated symbols, and the diversity gain that can be obtained by the transmitted bits varies between $1$ and $2$. In contrast, for the proposed CDD-AFDM-IM-I scheme, the transmitted bits consist of modulation bits and index bits. IM ensures that the index bits obtain a high diversity gain to enhance the reliability of the whole system. For example, see two different IM activation states ${{\bf{x}}_{i}}=[q_{i,1},~0,~0,~0]^T$ and ${{\bf{x}}_{j}}=[0,~q_{j,1},~0,~0]^T$.
Since the index bits are carried by the indices of the active chirp subcarriers, ${{\bf{x}}_{i}}-{{\bf{x}}_{j}}$ always has two non-zero elements.
Therefore, for any pair of ${{\bf{x}}_{i}}$ and ${{\bf{x}}_{j}}$ with different IM activation states in the CDD-AFDM-IM-I scheme, the rank of difference matrix ${\bf\Upsilon}_{{\bf{x}}_{i}-{\bf{x}}_{j}}$ is always $2$, i.e., ${\rm rank}\left({\bf\Upsilon}_{{\bf{x}}_{i}-{\bf{x}}_{j}}\right)=2$.
This is a simple proof, which shows that the IM scheme can obtain an additional diversity gain even the full diversity condition in~\eqref{CDD-AFDM.1} is not~satisfied.
\vspace{-4mm}
\subsection{Proposed CDD-AFDM-IM-II Scheme}
Based on the above analysis, we can observe that the diversity gain obtained by the index bits of the CDD-AFDM-IM-I scheme is limited by the number of non-zero elements in the difference vector ${{\bf{x}}_{i}}-{{\bf{x}}_{j}}$ if we further increase the value of $P{N}_{\rm t}$.
Therefore, to further enhance the transmission reliability, an effective approach is to increase the number of non-zero elements of ${{\bf{x}}_{i}}-{{\bf{x}}_{j}}$, thereby enhancing the potential to achieve a higher diversity order for the index bits in the IM scheme.
This motivates us to propose CDD-AFDM-IM-II scheme in which the chirp subcarrier activation states of the $g$ groups are the same.
For ease of exposition, we give an example of a pair DAF domain transmission symbols with different IM realizations as follows
\begin{equation}\label{CDD-AFDM.9}
{{\mathbf{x}}_{i}}\!=\!{{[\underbrace{0,~{{q}^{(1)}_{i,1}},~0,~0}_{1{\text -}{\rm th}~\rm{ group}},\underbrace{0,~{{q}^{(2)}_{i,1}},~0,~0}_{2{\text -}{\rm th}~\rm{ group}},\ldots ,\underbrace{0,~{{q}^{(g)}_{i,1}},~0,~0}_{g{\text -}{\rm th}~\rm{ group}}]}^{T}},
\end{equation}
\begin{equation}\label{CDD-AFDM.10}
{{\mathbf{x}}_{j}}\!=\!{{[\underbrace{{{q}^{(1)}_{j,1}},~0,~0,~0}_{1{\text -}{\rm th}~\rm{ group}},\underbrace{{{q}^{(2)}_{j,1}},~0,~0,~0}_{2{\text -}{\rm th}~\rm{ group}},\ldots ,\underbrace{{{q}^{(g)}_{j,1}},~0,~0,~0}_{g{\text -}{\rm th}~\rm{ group}}]}^{T}},
\end{equation}
where each group in ${{\mathbf{x}}_{j}}$ and ${{\mathbf{x}}_{j}}$ carriers the same index bits, and the number of index bits is $2$ in this example. It can be found that there are $2g$ non-zero elements in ${{\bf{x}}_{i}}-{{\bf{x}}_{j}}$.
The upper limit of the diversity order that can be achieved by the index bits in the CDD-AFDM-IM-II scheme is $g$ times higher than that in the CDD-AFDM-IM-I scheme.

We now present the realization of the proposed CDD-AFDM-IM-II scheme with high diversity.
For the CDD-AFDM-IM-II scheme, $N$ subsymbols in the DAF domain are divided into $L$ subblocks, and each
subblock is further divided into $g$-groups. Thus, the number of subsymbols in each group can be calculated as $n=N/gL$.
For the $\hat{l}$-$\rm{th}$ ($\hat{l}=1,\ldots,L$) subblock, the total transmitted bits is $p_{\hat{l}}=p/L$,
which consists of $p_{{\hat{l}},1}$ index bits and $p_{{\hat{l}},2}$ modulated bits.
The $p_{{\hat{l}},1}={\rm log}_2\left\lfloor C(n,m) \right\rfloor$ index bits are shared by $g$ groups to ensure the same IM activation state for each group.
The $p_{{\hat{l}},2}$ modulated bits are further divided into $g$ groups, and the $i$-$\rm{th}$ ($i=1,\ldots,g$) group needs to carry $p_{\hat{l},2}^{(i)}=p_{{\hat{l}},2}/g=m{\rm log}_2(M)$ modulated bits.
Then, based on the input $p_{{\hat{l}},1}$ index bits and $p_{\hat{l},2}^{(i)}$ modulated bits, the index selector and the mapper generate the index indication ${\bf k}_{\hat{l}}^{(i)}=\{{k}^{(i)}_{1};\ldots;{k}^{(i)}_{m}\}$ and constellation symbols ${\bf q}_{\hat{l}}^{(i)}=\{{q}^{(i)}_{1};\ldots;{q}^{(i)}_{m}\}$ of the $i$-$\rm{th}$ group, respectively.
Finally, according to ${\bf k}_{\hat{l}}=\{{\bf k}^{(1)}_{\hat{l}};\ldots;{\bf k}^{(g)}_{\hat{l}}\}$ and ${\bf q}=\{{\bf q}^{(1)}_{\hat{l}};\ldots;{\bf q}^{(g)}_{\hat{l}}\}$, we can obtain the transmitted symbol of the ${\hat{l}}$-$\rm th$ subblock as ${{\bf x}_{\hat{l}}=[{\bf x}_{\hat{l}}^{(1)};\ldots;{\bf x}_{\hat{l}}^{(g)}]}\in{\mathbb{C}}^{\frac{N}{L}\times1}$, and ${{\bf x}=[{\bf x}_{1};\ldots;{\bf x}_{L}]}\in{\mathbb{C}}^{{N}\times1}$.
The subsequent operations are performed as described in Section~\ref{section:SysMod} and the similar ABEP upper bound analysis can be discussed as~\eqref{23}.

{\bf{\em Remark~1:}} To realize high transmission rates and SEs, it is necessary to increase the number of subsymbols/chirp-carriers and the modulation order. Although we can use the ML detection directly to achieve the maximum diversity order with the optimal BER performance, the computational complexity of the ML detection grows exponentially with the number of subsymbols/chirp-carriers, i.e., the CDD-AFDM-IM-I as ${\mathcal{O}}(({{2}^{\frac{{p}_{1}}{g}}}M^{m})^{\frac{N}{n}})$, and the CDD-AFDM-IM-II as ${\mathcal{O}}(({{2}^{\frac{{p}_{1}}{gL}}}M^{m})^{\frac{N}{n}})$, where $p_1$ is the number of index bits that are carried by a CDD-AFDM-IM-I/II symbol.
This will hinder the application of the CDD-AFDM-IM-I/II schemes in practical communication systems. To tackle this problem, we develop an effective low-complexity detector for our proposed CDD-AFDM-IM-I/II schemes in the next section.
\section{Low Complexity Double-Layer Message Passing Detector}
\label{section:algorithm}
In this section, we develop the low-complexity DLMP algorithm for practical large-dimensional signal detection for our proposed CDD-AFDM-IM-I/II scheme.
We first describe the principle of the proposed DLMP detector and summarize its main steps in {\bf{Algorithm 1}}. We also discuss the corresponding computational complexity and make comparison with the existing benchmark detectors.
\begin{algorithm}[t]
  \caption{Proposed DLMP Detector}\label{DLMP-detector}
  \begin{algorithmic}[1]
   \STATE \textbf{Input}: $\bf y$, ${\bf{H}_{\rm eff}}$, ${N}_{0}$, and ${n}_{\rm{iter}}^{\rm{max}}$.\\
    \STATE \textbf{Initialization}:  Given parameters ${\rm{Pr}}_{c,r}=\frac{1}{\left|{\mathcal{M}_{B}}\right|}$, $\varsigma\in\{0,1\}$, ${f}_{c}(\varsigma)=\frac{1}{2}$, $c=1,2,\ldots,N$, ${\xi}^{(0)}=0$, and iteration count ${n}_{\rm{iter}}=1$.\\
     \STATE \textbf{Repeat}:\\
     \STATE Each observation node ${y}[r]$ computes the mean $\mu_{r,c}^{{{n}_{\rm iter}}}$ and variance~$\sigma_{r,c}^{{{n}_{\rm iter}}}$ of each connected variable node based on~\eqref{28} and~\eqref{29}. Then, each observation node ${y}[r]$ computes the probability estimate ${v}_{r,c}^{{n}_{\rm{iter}}}(x)$ based on~\eqref{30} and sends them to the variable node ${x}[c]$;\\
     \STATE Each indicator node ${a}[c]$ computes the posterior probability $f_c^{n_{\rm{iter }}}(\varsigma)$ of the activation and inactivation state for each variable node based on~\eqref{33.0} and sends them to the constraint node $G[i]$;\\
     \STATE Each constraint node generates ${u}_{c}^{n_{\rm{iter}}}(\varsigma)$ based on~\eqref{34} and passes ${u}_{c}^{n_{\rm{iter}}}(\varsigma)$ to the indicator node $a[c]$;\\
     \STATE Each variable node ${x}[c]$ computes the extrinsic probability ${\bf{{Pr}}}_{c,r}^{{{n}_{{\rm iter}}}}$ based on~\eqref{31} and passes them to the connected observation nodes ${y}[r]$;\\
     \STATE Compute the convergence indicator ${\xi}^{n_{\rm{iter}}}$ and probability of the transmitted symbols ${\bf{{Pr}}}_{c}^{{{n}_{{\rm iter}}}}$ based on~\eqref{35} and~\eqref{36}, respectively;\\
     \STATE if ${\xi}^{n_{\rm{iter}}}\!>\!{\xi}^{(n_{\rm{iter}}-1)}$, update the probabilities ${\bf{\overline{Pr}}}_{c}={\bf{{Pr}}}_{c}^{{n}_{\rm{iter}}}$;\\
     \STATE ${n}_{\rm{iter}}={n}_{\rm{iter}}+1$;\\
     \STATE \textbf{Until}: ${\xi}^{n_{\rm{iter}}}=1$ or ${n}_{\rm iter}={n}_{\rm iter}^{\rm max}$.\\
     \STATE {\textbf{Output}: The estimated activation state of the transmitted symbols and the estimated transmitted symbols $\widehat{\mathbf{x}}$.}
  \end{algorithmic}
\end{algorithm}
\subsection{Proposed DLMP Detector}
\begin{figure}[t]
	\center
	\includegraphics[width=2.8in,height=2.0in]{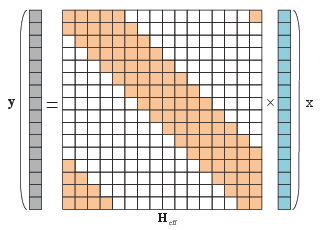}
	\vspace{-0.2cm}
	\caption{Input-output relation between ${\bf x}$ and ${\bf y}$, where $N=16$, ${N}_{\rm t}=2$, ${l}_{\rm{max}}=0$, and ${\alpha}_{\rm{max}}=1$.}
	\label{Structure-H}  
	\vspace{-2mm}
\end{figure}

\begin{figure}[t]
	\center
	\includegraphics[width=3.2in,height=2.0in]{{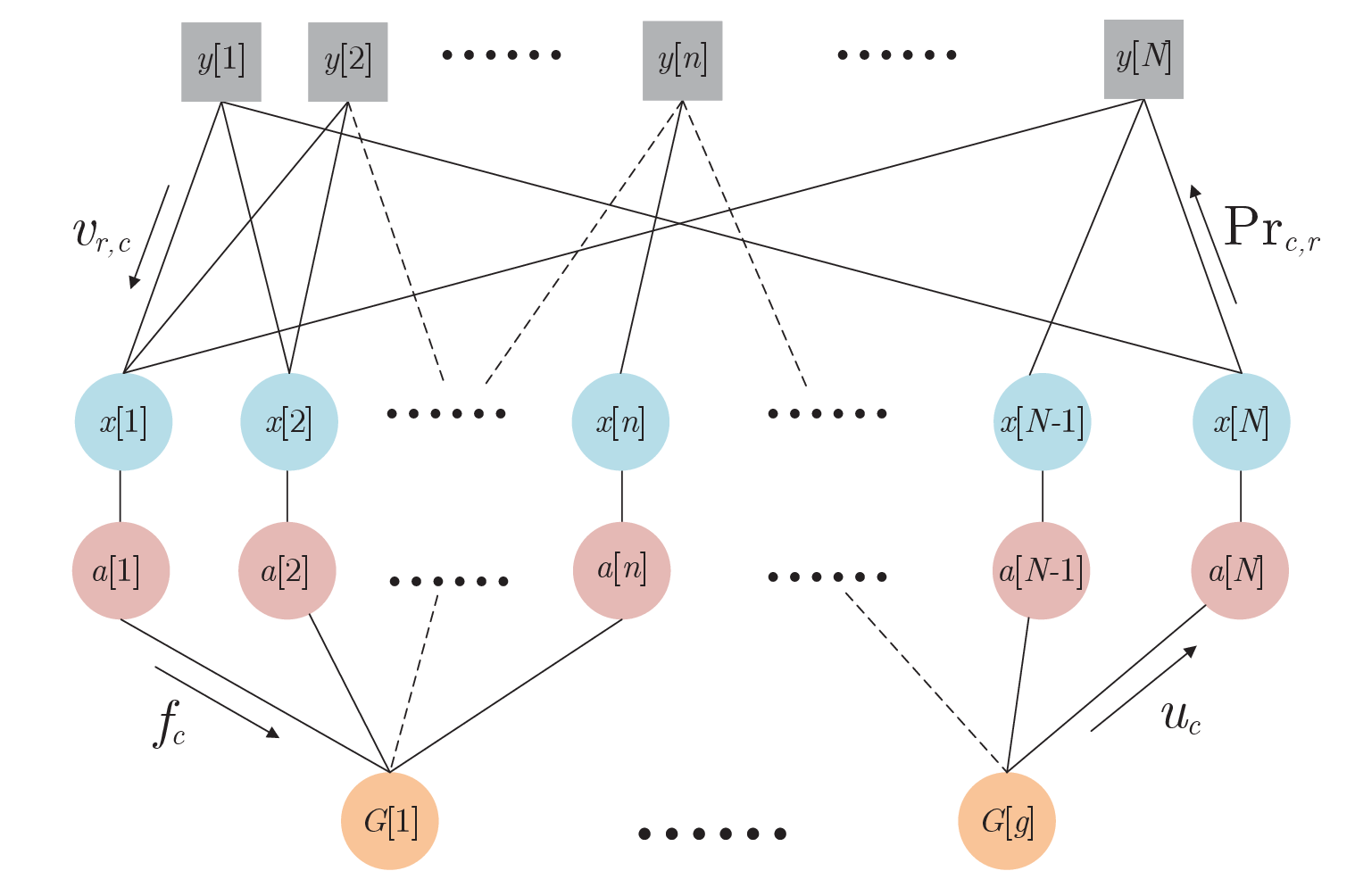}}
	\vspace{-0.2cm}
	\caption{Graphical model and the messages passed in the proposed DLMP detector.}
	\label{Graphical-model}  
	\vspace{-6mm}
\end{figure}

According to~\eqref{8}, we can get an example of the input-output relation of the proposed CDD-AFDM-IM-I/II schemes as shown in Fig.~\ref{Structure-H}. The prefect CSIs are assumed at the receiver. Therefore, the expression of joint maximum a-posteriori probability (MAP) detection rule for the transmitted signal is given by
\begin{equation}\label{24}
\widehat{\bf x}= \underset{{\bf x}\in{{\mathcal{M}}_{B}^{N\times1}}}{\rm{arg~max }}\,{\rm{Pr}}{\left({\bf{x}|{\bf{y}},{\bf{H}_{\rm{eff}}}}\right)},
\end{equation}
where ${\mathcal{M}}_{B}=\{{\mathcal{M}}_{A}\cup0\}$, ${\mathcal{M}}_{A}$ is the set of constellations for the modulated bits mapping, and `0' indicates that the subsymbol is in an inactive state.
The exact computation of~\eqref{24} has a complexity exponential in $N$, making the joint MAP detection intractable for practical large values of $N$.
As a result, we give a suboptimal MAP detection with low complexity, and
${\rm{Pr}}{\left({\bf{x}|{\bf{y}},{\bf{H}}_{\rm{eff}}}\right)}$ can be rewritten as{\footnote{Similar to the expression of the CDD-AFDM-IM-I scheme, the expression of the CDD-AFDM-IM-II scheme is the same when $L=1$, and the proposed DLMP algorithm can be easily generalized to the case of $L>1$. Therefore, in the following presentation, we assume that $L=1$ to clarify the principle of the proposed DLMP algorithm more clearly.}}
\begin{align}\label{25}
  & \Pr \left({\bf x},{\bf a}|{\bf y},{\bf H}_{\rm{eff}} \right) \nonumber\\
 & \propto \Pr \left( {\bf y}|{\bf x},{\bf a},{\bf H}_{\rm{eff}} \right)\Pr \left({\bf x},{\bf a} \right) \nonumber\\
 & =\Pr \left({\bf y}|{\bf x},{\bf a},{\bf H}_{\rm{eff}} \right)\Pr \left({\bf x}|{\bf a} \right)\Pr \left({\bf a} \right)\nonumber\\
 & =\left\{\!\prod\limits_{r=1}^{N}\!{\Pr \left({y}[r]|{\bf x},{\bf a},{\bf H}_{\rm{eff}} \right)}\!\!\!\!\!\!\!\prod\limits_{c=(i-1)n+1}^{in}\!\!\!\!\!\!\!\!\!{\Pr \left({x}[c]|{a}[c] \right)} \!\right\}\!\Pr \left({\bf a} \right),
\end{align}
where ${\bf a}=[a[1],a[2],\ldots,a[gn]]^{T}\in{{\mathbb{C}}^{N\times1}}$ represents the vector of the activation state of all subsymbols.
We give the graphical model and message passing process of the proposed DLMP algorithm in Fig.~\ref{Graphical-model}.
The graphical model of the DLMP algorithm consists of four types of nodes, i.e.,
$N$ observation nodes and $N$ variable nodes corresponding to the elements in ${\bf{y}}$ and ${\bf{x}}$, respectively, $N$ indicator nodes corresponding to the activation state of $N$ subsymbols, and $g$ constraint nodes.
Based on the input-output relation shown in Fig.~\ref{Structure-H}, we can find that the connections between the observation nodes and the variable nodes are determined by the channel matrix ${\bf H}_{\rm eff}$.
Let us define ${\mathcal{K}}[c]$ and ${\mathcal{L}}[r]$ to be the indices of the non-zero elements in the $c$-column and $r$-row of ${\bf H}_{\rm eff}$.
Hence, each observation node ${{y}}[r]$ connects ${N}_{\rm{t}}{P}_{\rm{max}}$ variable nodes ${{x}}[c], c\in{\mathcal{L}}[r]$.
Similarly, each variable node ${{x}}[c]$ is associated with ${N}_{\rm{t}}{P}_{\rm{max}}$ observation nodes ${{y}}[r], r\in{\mathcal{K}}[c]$.
Moreover, the relationship between a variable node and an indicator node can be expressed as
\begin{align}\label{26}
{a}[c]=\left\{ \begin{matrix}
   1, &~~~~~~~{x}[c]\in {\mathcal M}_{A},  \\
   0, &~~~~~~~{\rm otherwise.}  \\
\end{matrix} \right.
\end{align}
Since each group consists of $n$ subsymbols in the DAF domain, each constraint node ${{G}}[i]$ connects $n$ corresponding nodes ${{a}}[c]$, where $c\in\{(i-1)n+1,\ldots,in\}$.
In the proposed DLMP algorithm, the observation nodes and the variable nodes constitute the first layer, which is used to calculate the posterior probability of each subsymbol with respect to ${\mathcal{M}}_{B}$.
Indicator nodes and constraint nodes constitute the second layer, which generates the probability of subsymbols activation/inactivation in the DAF domain based on the posteriori probabilities output from the first layer.
The detailed steps and iterative process of the proposed DLMP algorithm are described as below.
\subsubsection{\bf\em{Observation and variable nodes (Layer~1)}}
First, at each observation node, we need to compute the extrinsic information for each connected variable node based on the channel model, observation noise, and a priori information from other connected variable nodes.
According to~\eqref{8}, we can obtain the expression of the observation node ${y}[r]$ as
\begin{align}\label{27}
{y}[r]={{H}_{\rm eff}}[r,c]{x}[c]+\underbrace{\sum\limits_{e\in {\mathcal{L}}[r],e\ne c}{{{H}_{\rm eff}}[r,e]}{x}[e]+\overline{w}[r]}_{\wp_{r,c}}.
\vspace{-2mm}
\end{align}
We use a Gaussian approximation of the interference term to obtain the approximate marginal probabilities of the transmitted symbols. According to the central limit theorem (CLT), ${\wp_{r,c}}$ in~\eqref{27} obeys the Gaussian distribution with mean $\mu_{r,c}$ and variance ${{\left( \sigma _{r,c} \right)}^{2}}$. The mean and variance of the ${n}_{\rm iter}$-$\rm th$ iteration can be given by
\begin{align}\label{28}
\mu_{r,c}^{{{n}_{\rm iter}}}=\sum\limits_{e\in {\mathcal{L}}[r],e\ne c}{{{H}_{\rm eff}}[r,e]}\sum\limits_{x\in {{\mathcal{M}}_{B}}}{x}{{\rm Pr}_{e,r}^{\left( {{n}_{\rm{iter}}}-1 \right)}(x)},
\vspace{-2mm}
\end{align}
and
\begin{align}\label{29}
   {{\left( \sigma _{r,c}^{{{n}_{\rm{iter}}}} \right)}^{2}}=&\sum\limits_{e\in \mathcal{L}[r],e\ne c}{{{\left| {{H}_{\rm{eff}}}[r,e] \right|}^{2}}}\left( \sum\limits_{x\in {{\mathcal{M}}_{B}}}{{{\left| x \right|}^{2}}}{\rm{Pr}}_{e,r}^{\left( {{n}_{\rm{iter}}}-1 \right)}(x) \right. \nonumber\\
 & \left. -{{\left| \sum\limits_{x\in {{\mathcal{M}}_{B}}}{x}{\rm{Pr}}_{e,r}^{\left( {{n}_{\rm{iter}}}-1 \right)}(x) \right|}^{2}} \right)+{{N}_{0}},
 \vspace{-2mm}
\end{align}
where ${\rm{Pr}}_{e,r}(x)$ denotes the extrinsic probability of the symbol ${x}$ passed from variable node $x[e]$ to observation node $y[r]$.
At the observation nodes, based on the obtained mean ${\mu_{r,c}^{{n}_{\rm iter}}}$ and variable $\left({\sigma_{r,c}^{{n}_{\rm iter}}}\right)^{2}$, 
we calculate the probability estimate of the elements $x[c]$ as follows, which will be passed from observation node $y[r]$ to variable node $x[c]$
\begin{align}\label{30}
{v}_{r,c}^{{n}_{\rm{iter}}}(x)&\propto{{\rm{Pr}}\left({{x}[c]=x}|{y[r]},{\bf{H}}_{\rm{eff}}\right)}\nonumber\\
&\propto
{\rm{exp}}\left(\frac{-\left|{{y}[r]}-\mu_{r,c}^{{{n}_{\rm iter}}}-{H_{\rm{eff}}[r,c]}x\right|^2}{{{\left( \sigma _{r,c}^{{{n}_{\rm{iter}}}} \right)}^{2}}}
\right).
\end{align}
Moreover,
according to the a priori information passed from the indicator node ${a}[c]$ and observation nodes $y[r]$, $r\in{{\mathcal{K}[c]}}$ to the variable nodes ${x}[c]$,
the extrinsic probability ${\rm{Pr}}_{c,r}(x)$ passed from a variable node ${x}[c]$ to the observation nodes $y[r]$ can be expressed as a probability mass function (PMF)
\begin{align}\label{31}
{\rm{{Pr}}}_{c,r}^{{{n}_{{\rm iter}}}}(x)={\varpi}{\rm{\widetilde{Pr}}}_{c,r}^{{{n}_{\rm iter}}}(x)+(1-{\varpi}){\rm{{Pr}}}_{c,r}^{{{n}_{{\rm iter}-1}}}(x),
\end{align}
where
\begin{align}\label{32}
   {\rm{\widetilde{Pr}}}_{c,r}^{{{n}_{\rm iter}}}(x)&\propto {\rm{Pr}}(x[c]=x|{{\bf y}_{\backslash r}},{\bf H}_{\rm{eff}}) \nonumber\\
 & \propto u_{c}^{{{n}_{\rm iter}}}(\varsigma)\!\!\!\prod\limits_{e\in {\mathcal{K}}[c],e\ne r}\!\!\!{{\rm{Pr}}(x[c]=x|y[e],{\bf H}_{\rm{eff}})} \nonumber\\
 & \propto u_{c}^{{{n}_{\rm iter}}}(\varsigma)\!\!\!\prod\limits_{e\in {\mathcal{K}}[c],e\ne r}\!\!\!{v_{e,c}^{{{n}_{\rm iter}}}(x)},~~~\forall x\in{{\mathcal{M}}_{B}},
\end{align}
with $\varsigma=1$ if $x\in{{\mathcal{M}}_{A}}$, otherwise $\varsigma=0$, and $u_{c}^{{{n}_{\rm iter}}}(\varsigma)$ being the priori information passed from the second layer to the variable node $x[c]$.
The message damping factor $\varpi\in(0,1]$ is used to control the convergence rate of the algorithm and improve stability~\cite{9349154,9354639}.
\subsubsection{\bf\em{Indicator and constraint nodes (Layer~2)}}
In the second layer of the algorithm, the indicator nodes calculate the posterior probability the activation and inactivation state for each variable node, which is given by
\begin{align}\label{33.0}
 f_c^{n_{\rm{iter }}}(\varsigma)={\varpi} {\widetilde{f}}_c^{n_{\rm{iter }}}(\varsigma)+(1-{\varpi}) f_c^{n_{\rm{iter-1 }}}(\varsigma),
\end{align}
where
\begin{align}\label{33}
 {\widetilde{f}}_c^{n_{\rm{iter }}}(\varsigma) &\triangleq \operatorname{Pr}(a[c]=\varsigma \mid \mathbf{x}) \nonumber\\
& \propto\left\{\begin{array}{lll}
\sum\limits_{x \in \mathcal{M}_A} \prod\limits_{r \in \mathcal{K}[c]} \operatorname{Pr}(x[c]=x \mid y[r], \mathbf{H}_{\rm{eff}}), & {\rm if } & \varsigma=1, \\
\prod\limits_{r \in \mathcal{K}[c]} \operatorname{Pr}(x[c]=0 \mid y[r], \mathbf{H}_{\rm{eff}}), &{\rm if } & \varsigma=0,
\end{array}\right. \nonumber\\
& \propto \begin{cases}\sum\limits_{x \in \mathcal{M}_A} \prod\limits_{r \in \mathcal{K}[c]} {v}_{r,c}^{{n}_{\rm{iter}}}(x), &  {\rm if } \quad \varsigma=1, \\
\prod\limits_{r \in \mathcal{K}[c]} {v}_{r,c}^{{n}_{\rm{iter}}}(0), & {\rm if } \quad \varsigma=0.\end{cases}
\end{align}
Then, ${{f}}_c^{n_{\rm{iter }}}(\varsigma)$ is regraded as a priori information passed from the indicator node $a[c]$ to the constraint node $G[i]$, where ${c\in{\mathcal{U}}[i]}$, and ${\mathcal{U}}[i]=[(i-1)n+1,\ldots,in]$.
Since a constraint node corresponds to $m$ activated indicator nodes, the condition of the constraint node ${{G}}[i]$ can be expressed as $\sum\nolimits_{c\in{\mathcal{U}}[i]}{{a}}[c]=m$.
After processing by the constraint node, the message $u_{c}$ is passed from the constraint node $G[i]$ to the indicator node $a[c]$, which can be calculated as
\begin{align}\label{34}
{u}_{c}^{n_{\rm{iter}}}(\varsigma)&={\rm{Pr}}\left(a[c]=\varsigma|{{\bf{a}}_{\backslash c}^{(i)}}\right)\nonumber\\
 & \propto \left\{ \begin{array}{*{35}{l}}
   \Pr \left( \sum\limits_{e\in{{\mathcal U }[i]},e\ne c}{a[e]=m-1}\mid \mathbf{a}_{\backslash c}^{(i)} \right), &{\rm if} & \varsigma=1,  \\
   \Pr \left( \sum\limits_{e\in{{\mathcal U}[i]},e\ne c}{a[e]=m}\mid \mathbf{a}_{\backslash c}^{(i)} \right), & {\rm if} & \varsigma=0,  \nonumber\\
\end{array} \right. \\
 & \propto \left\{ \begin{array}{*{35}{l}}
   \Im_{c}^{n_{\rm{iter}}} (m-1), & {\rm if}\quad \varsigma=1,  \\
   \Im_{c}^{n_{\rm{iter}}} (m), & {\rm if}\quad \varsigma=0,  \\
\end{array} \right.
\end{align}
where $\mathbf{a}^{(i)}=[a[(i-1)n+1],\ldots,a[in]]$, $\mathbf{a}_{\backslash c}^{(i)}\cup{a[c]}={\mathbf{a}^{(i)}}$, and $c\in{\mathcal{U}}[i]$.
$\Im_{c}^{n_{\rm{iter}}}$ can be calculated as $\Im_{c}^{n_{\rm{iter}}}={{\otimes }_{e\in{\mathcal{U}[i]},e\ne c }}{\bf{f}}_{e}^{n_{\rm iter}}$ with ${\bf{f}}_{e}^{n_{\rm iter}}=[{{f}}_{e}^{n_{\rm iter}}(0),{{f}}_{e}^{n_{\rm iter}}(1)]$.
\subsubsection{\bf\em{Convergence indicator}} After a sufficient number of message iterations, the convergence indicator can be calculated as
\begin{align}\label{35}
{\xi}^{n_{\rm{iter}}}=\frac{1}{N}\sum\limits_{c=1}^{N}\mathbb{I}
\left(
\underset{x\in{\mathcal{M}_{B}}}{\rm max} {\rm Pr}_{c}^{n_{\rm{iter}}}(x)\ge1-{\mathcal P}
\right)
\end{align}
for some small ${\mathcal P}>0$, and $\mathbb{I}(\cdot)$ is the indicator function. The estimate MAP of each element of the transmit symbol $\bf x$ is
\begin{align}\label{36}
   {\rm Pr}_{c}^{n_{\rm{iter}}}(x)=\frac{1}{C}
   u_{c}^{{{n}_{\rm iter}}}(\varsigma)\prod\limits_{r\in {\mathcal{K}}[c]}{v_{r,c}^{{{n}_{\rm iter}}}(x)},~~~\forall x\in{{\mathcal{M}}_{B}},
\end{align}
where $C$ is the normalizing constant.
\subsubsection{\bf\em{Update criteria}}
We only update the probabilities if the current iteration convergence indicator is better than the previous one. This criteria can be represented as
\begin{align}\label{37}
  {\bf{\overline{Pr}}}_{c}={\bf{{Pr}}}_{c}^{{n}_{\rm{iter}}}, &~~~~~{\rm if}\quad {\xi}^{n_{\rm{iter}}}>{\xi}^{(n_{\rm{iter}}-1)}.
\end{align}

\subsubsection{\bf\em{Stopping criteria}}
The DLMP algorithm stops when the output ${\xi}^{n_{\rm{iter}}}$ of the convergence indicator is $1$ or the number of maximum iterations ${n}_{\rm iter}^{\rm max}$ is reached.
Afterwards, we determine the activation state for each group based on the posterior probability $f_c^{n_{\rm{iter }}}(\varsigma)$.
Specifically, for the $i$-$\rm{th}$ group, we can obtain an activation probability vector for subsymbols as ${\bf{f}}^{(i)}=[f_{(i-1)n+1}^{n_{\rm{iter }}}(1);\ldots;f_{\beta}^{n_{\rm{iter }}}(1);\ldots;f_{in}^{n_{\rm{iter }}}(1)]$, where $\beta\in{\mathcal{U}}[i]$.
For the CDD-AFDM-IM-I scheme, we can find the $m$ largest probabilities from the activation probability vector ${\bf{f}}^{(i)}$. Based on the indicator nodes corresponding to the $m$ largest probabilities, we can determine the activation state of subsymbols in the $i$-$\rm{th}$ group.
Different from the CDD-AFDM-IM-I scheme, the proposed CDD-AFDM-IM-II scheme uses the same activation state of subsymbols for $g$ groups.
Therefore, we merge the activation probability vectors of the $g$ groups to obtain the final activation probability vector, i.e., ${\bf{f}}=\left[\sum\nolimits_{i=1}^{g}f_{(i-1)n+1}^{n_{\rm{iter }}}(1);\ldots;\sum\nolimits_{i=1}^{g}f_{in}^{n_{\rm{iter }}}(1)\right]$.
Similarly, $m$ largest activation probabilities are found from the probability vector $\bf{f}$.
The subsymbol activation states of the $g$ groups are obtained based on the indicator nodes corresponding to the $m$ largest probabilities.
Subsequently, we can perform the estimation of the modulated symbols:
\begin{equation}\label{39}
{\hat{x}}[c]= \underset{x\in{\mathcal{M}_{A}}}{\rm{arg~max}}~{{\overline{\rm Pr}}}_{c}(x),~c\in{\mathcal{R}_{a}},
\end{equation}
where ${\mathcal{R}_{a}}$ is the indices of the position for the activation subsymbols.
Finally, based on the estimated symbols and the activation state of each group, we can easily obtain the transmitted bits by demapping.
\vspace{-3mm}
\begin{figure*}[ht]
\begin{align}\label{40}
  {{\mathcal{O}}_{\rm{DLMP}}} =&\left(\underbrace{P{{N}_{\rm{t}}}\left( 4{{M}}+10 \right)N-2N}_{\eqref{28}}\right.+\underbrace{P{{N}_{\rm{t}}}(9{{M}}+15)N}_{\eqref{29}}
  +\underbrace{17P{{N}_{\rm{t}}}\left( {{M}}+1 \right)N}_{\eqref{30}}+\underbrace{P{{N}_{\rm{t}}}\left( {{M}}+1 \right)N}_{\eqref{32}}\nonumber\\
 & +\underbrace{P{{N}_{\rm{t}}}\left( {{M}}+1 \right)N-2N}_{\eqref{33}}
   \left.+\underbrace{\left[ \frac{3\left( n-1 \right)\left( n-2 \right)}{2}-3 \right]N}_{\eqref{34}}\right)n_{\rm iter}^{\rm ave}
\end{align}
\hrulefill
\vspace{-4mm}
\end{figure*}
\subsection{Computational Complexity Analysis}
The complexity of the proposed DLMP detector mainly depends on the iterative process of the DLMP algorithm.
We quantify the computational complexity based on the real-valued floating point operations (FLOPs) during the iterations. We consider a complex multiplication and addition to correspond to six and two real-valued floats, respectively. Also, a real-valued multiplication or addition corresponds to one real-valued floating point~\cite{8039493,7458894}.
In the DLMP algorithm, the computational complexity of each iteration is dominated by~\eqref{28}$-$\eqref{30},~\eqref{32},~\eqref{33} and~\eqref{34}.
We define $n_{\rm iter}^{\rm ave}$ as the average number of iterations required for convergence. Thus, the overall computational complexity of the the DLMP algorithm can be calculated as~\eqref{40}, shown at the top of this page.

Similarly, we can compute the computational complexity of the MMSE algorithm~\cite{tao2023affine} and the conventional single-layer MP algorithm~\cite{8424569} as
${{\mathcal{O}}_{\rm{MMSE}}}=16{N}^3+13{N}^2$ and ${{\mathcal{O}}_{\rm{MP}}}=\left(NP{{N}_{t}}\left( 31{{M}}+43 \right)-2N\right)n_{\rm iter}^{\rm ave}$, respectively. We can observe that the proposed DLMP algorithm enjoys lower computational complexity than ML and MMSE algorithms.
Although the computational complexity of the proposed DLMP algorithm is slightly higher than that of the single-layer MP detection, the proposed DLMP algorithm achieves a better BER performance, leading to an attractive tradeoff between BER performance and computational complexity. We will discuss this issue in detail in the next section by simulation results.
\section{Simulation Results and Discussions}
\label{section:Simulation}
In this section, we first evaluate the BER performance of the proposed CDD-AFDM-IM-I/II schemes by using the ML detection and verify the accuracy of our theoretical derivation. Then, the complexity and BER performance of the proposed low-complexity DLMP algorithm for the large-scale CDD-AFDM-IM-I/II schemes are investigated.
Without loss of generality, we set the carrier frequency as $4~{\rm{GHz}}$, the subsymbol spacing in the DAF domain as $15~{\rm{kHz}}$, and the modulation alphabet as BPSK/QPSK.
Unless otherwise stated,
we set the maximum Doppler as ${\alpha}_{\rm{max}}=1$ corresponding to the maximum speed of $405~{\rm{km/h}}$, and the number of maximum multipaths as ${P}_{\rm max}=3$.
We use the Jakes Doppler spectrum approach to generate the Doppler shifts, i.e., $\alpha_{\epsilon,\ell}={\alpha_{\rm max}{\rm{cos}}(\theta_{\epsilon,\ell})}$, where $\theta_{\epsilon,\ell}\in[-\pi,\pi]$ obeys a uniform distribution~\cite{9293173}.
\vspace{-3mm}
\begin{figure}[t]
	\center
	\includegraphics[width=2.8in,height=2.0in]{{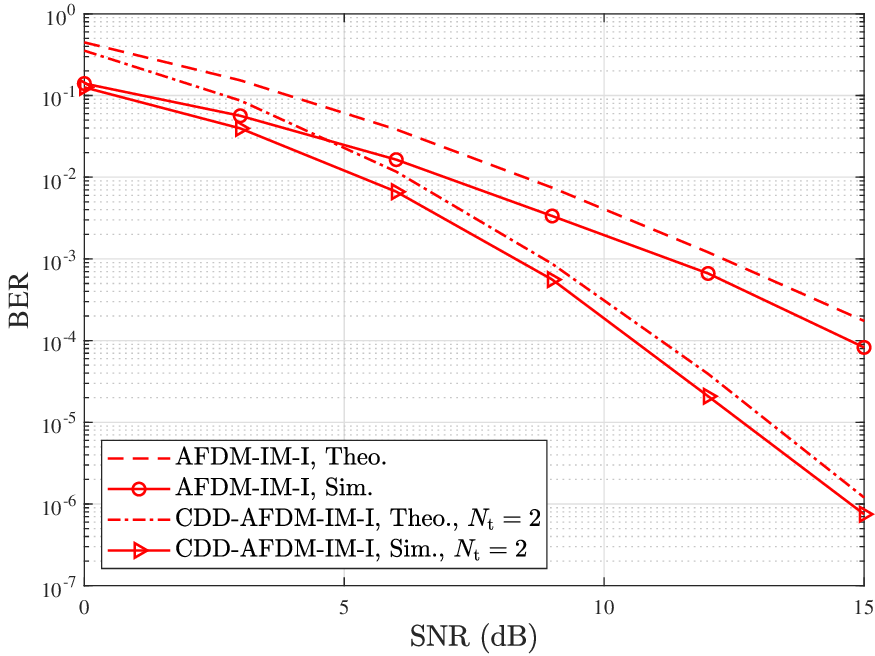}}
	\vspace{-0.1cm}
	\caption{Simulated and theoretical BER performance of the proposed CDD-AFDM-IM-I scheme with BPSK over an LTV channel, where $N=10$, $g=1$, $m=1$, $P=3$, and ${N}_{\rm t}=1,2$.}
	\label{Sim-Theo}  
	\vspace{-0mm}
\end{figure}
\begin{figure}[t]
	\center
	\includegraphics[width=2.8in,height=2.0in]{{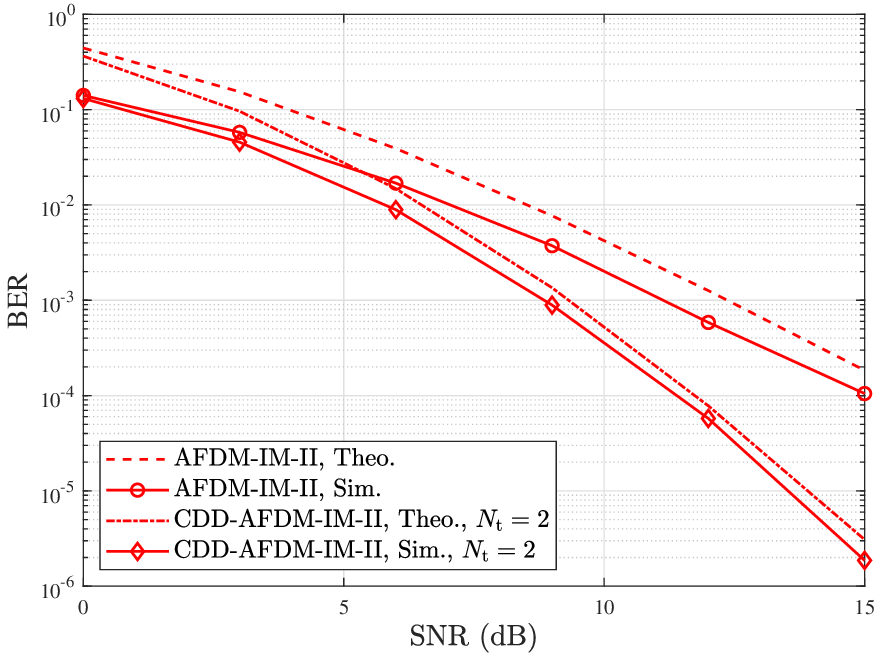}}
	\vspace{-0.1cm}
	\caption{Simulated and theoretical BER performance of the proposed CDD-AFDM-IM-II scheme with BPSK over an LTV channel, where $N=8$, $g=2$, $m=1$, $P=3$, and ${N}_{\rm t}=1,2$.}
	\label{Sim-Theo-II}  
	\vspace{-0mm}
\end{figure}
\subsection{BER Performance by Using ML Detector}
Figs.~\ref{Sim-Theo} and~\ref{Sim-Theo-II} illustrate the simulated and theoretical BER performance of the proposed CDD-AFDM-IM-I/II schemes over the LTV channnel.
Note that the CDD-AFDM-IM-I/II schemes degenerate to AFDM-IM-I/II scheme when ${N}_{\rm{t}}=1$.
One can observe that the simulated and theoretical curves match well in the high SNR region, which indicates the validity of the upper bound analysis in~\eqref{23}.
In addition, one can notice that the BER performances of the proposed CDD-AFDM-IM-I/II schemes improve significantly as the number of TAs increases.
This is due to the fact that the proposed CDD-AFDM-IM-I/II framework can achieve the transmit spatial diversity by performing a cyclic delay operation to the transmitted signals for each TA.
The diversity gains of the proposed CDD-AFDM-IM-I/II schemes increase with the increase of ${N}_{\rm{t}}$.

\begin{figure}[t]
	\center
	\includegraphics[width=2.8in,height=2.0in]{{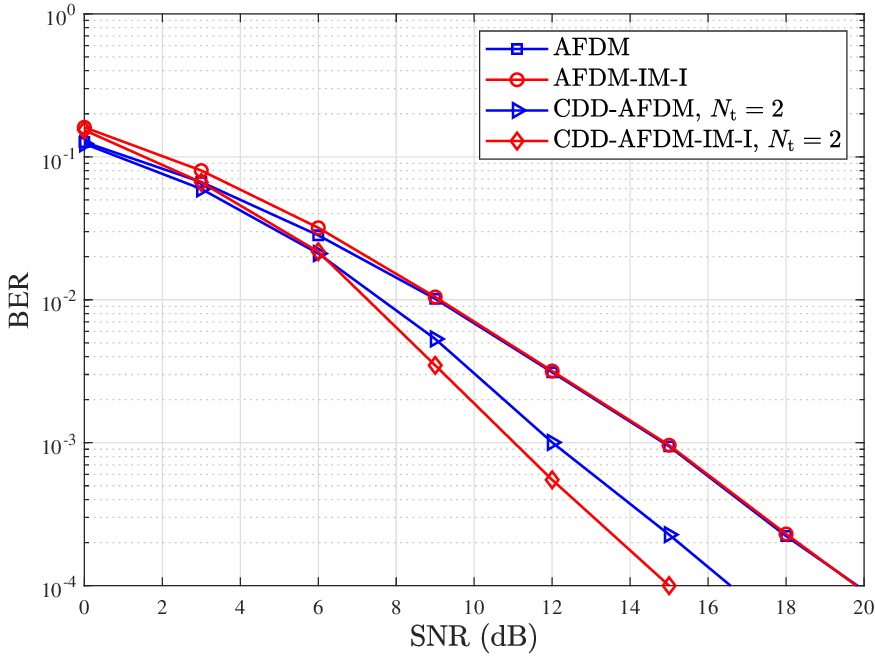}}
	\vspace{-0.2cm}
	\caption{BER performance of the proposed CDD-AFDM-IM-I, AFDM-IM-I, CDD-AFDM, and AFDM schemes by using the ML detection for the integer Doppler case over an LTV channel, where $N=4$, $g=1$, $m=1$, $P=2$, and ${N}_{\rm t}=1,2$.}
	\label{int-ML-com-I}  
	\vspace{-2mm}
\end{figure}
\begin{figure}[t]
	\center
	\includegraphics[width=2.8in,height=2.0in]{{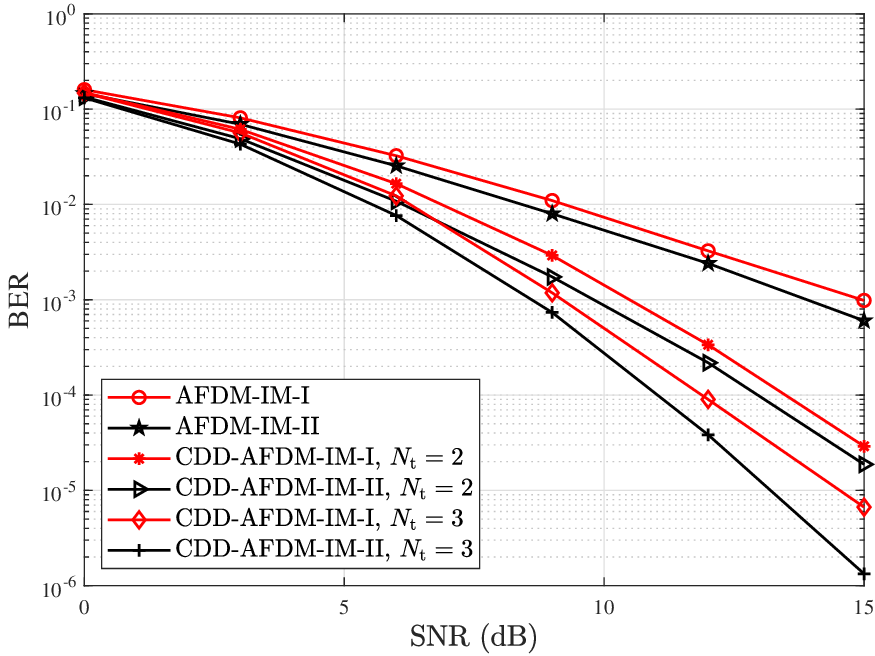}}
	\vspace{-0.2cm}
	\caption{BER performance of the proposed CDD-AFDM-IM-II, CDD-AFDM-IM-I, AFDM-IM-II, and AFDM-IM-I schemes by using the ML detection for the integer Doppler case over an LTV channel, where $N=8$, $g=2$, $m=1$, $P=3$, and ${N}_{\rm t}=1,2,3$.}
	\label{int-ML-com-II}  
	\vspace{-6mm}
\end{figure}
\begin{figure}[t]
        \centering
        \subfigure[\hspace{-0.2cm}]{\label{int-MP-com}
        \includegraphics[width=2.8in,height=2.0in]{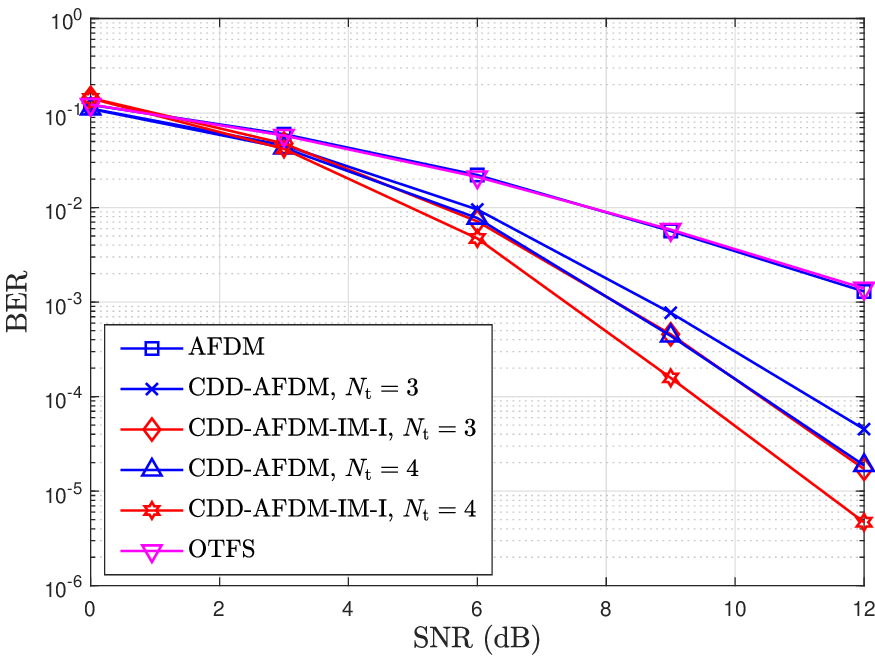}}\vspace{0.3cm}
        \subfigure[\hspace{-0.2cm}]{\label{fra-MP-com}
        \includegraphics[width=2.8in,height=2.0in]{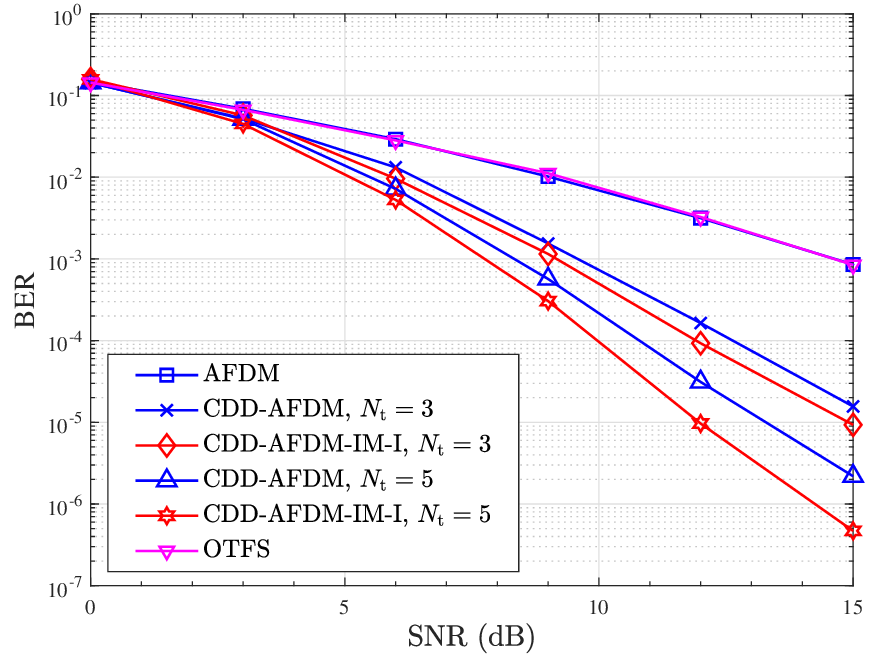}}
        \vspace{-0.0cm}
    	\caption{BER performance of the proposed CDD-AFDM-IM-I, AFDM, CDD-AFDM, and OTFS schemes by using the DLMP detection for (a) integer Doppler and (b) fractional Doppler cases over an LTV channel.
    }
    	\label{int-fra-MP-com-I}  
    	\vspace{-2mm}
\end{figure}

In Fig.~\ref{int-ML-com-I}, we investigate the BER performance of the proposed CDD-AFDM-IM-I, AFDM-IM-I, CDD-AFDM, and AFDM schemes when the full diversity condition in~\eqref{CDD-AFDM.1} is satisfied/not satisfied.
We apply the QPSK mapping for the modulated bits in CDD-AFDM-IM-I and  AFDM-IM-I schemes, and adopt the BPSK mapping in AFDM and CDD-AFDM schemes, leading to the same SE (i.e., $1~\rm{bits/Hz}$) for different schemes.
One can observe that the transmission framework satisfies the full diversity condition (i.e., ${P}_{\rm{max}}<N$) when the number of TAs is 1.
In this case, one can notice from Fig.~\ref{int-ML-com-I} that the AFDM scheme and the AFDM-IM-I scheme experience similar BER performance especially in the high SNR region.
This is because the diversity gains that can be obtained by the AFDM-IM-I scheme and AFDM are similar when the full diversity condition is satisfied.
Moreover, when ${N}_{\rm{t}}=2$, the full diversity conditions of the CDD-AFDM-IM-I scheme and the CDD-AFDM scheme in~\eqref{CDD-AFDM.1} are no longer satisfied. In this case, one can notice that the CDD-AFDM-IM-I scheme significantly outperforms the CDD-AFDM scheme.
This is expected because when the full diversity condition in~\eqref{CDD-AFDM.1} is not satisfied, the IM-assisted scheme can provide additional diversity to achieve stronger diversity protection and better BER performance.
This result is also consistent with the diversity analysis in Section~\ref{section:IM.A}.

Fig.~\ref{int-ML-com-II} further illustrates the BER performance of the proposed CDD-AFDM-IM-I and AFDM-IM-I schemes and compares them to their corresponding improved versions, i.e., CDD-AFDM-IM-II and AFDM-IM-II schemes.
Similarly, we apply the QPSK mapping for the modulated bits in CDD-AFDM-IM-II and  AFDM-IM-II schemes, and adopt the BPSK mapping in AFDM-IM-I and CDD-AFDM-IM-I schemes, leading to the same SE (i.e., $0.75~\rm{bits/Hz}$) for different schemes.
One can observe that the BER performance of the proposed CDD-AFDM-IM-II and AFDM-IM-II schemes outperforms that of the CDD-AFDM-I and AFDM-I schemes.
We also notice that the performance gap between the CDD-AFDM-IM-II scheme and the CDD-AFDM-IM-I scheme are similar when the full diversity condition in~\eqref{CDD-AFDM.1} is satisfied (i.e., the number of TAs ${N}_{\rm t}$ is 1 or 2).
In contrary, when the number of TAs increases to the level that the full diversity condition in~\eqref{CDD-AFDM.1} is broken (i.e., ${N}_{\rm t}=3$), the BER performance gain of the CDD-AFDM-IM-II scheme will be further improved in comparison with the CDD-AFDM-IM-I scheme.
This is because under the full diversity condition in~\eqref{CDD-AFDM.1}, the CDD-AFDM-IM-I scheme and the {CDD-AFDM-IM-II} scheme are able to obtain the same diversity order, and the performance gain comes from the fact that the $g$ groups in the CDD-AFDM-IM-II scheme take the same activation sate to jointly transmit the index bits, thus improving the performance of index bits.
Correspondingly, when the full diversity condition in~\eqref{CDD-AFDM.1} is not satisfied, the CDD-AFDM-IM-II scheme can achieve a higher diversity order than the CDD-AFDM-IM-I scheme, which can be verified in the diversity analysis in Section~\ref{section:IM}.
\vspace{-3mm}
\subsection{BER Performance by Using the Low Complexity Detector}
\begin{figure}[t]
	\center{
	\includegraphics[width=2.8in,height=2.0in]{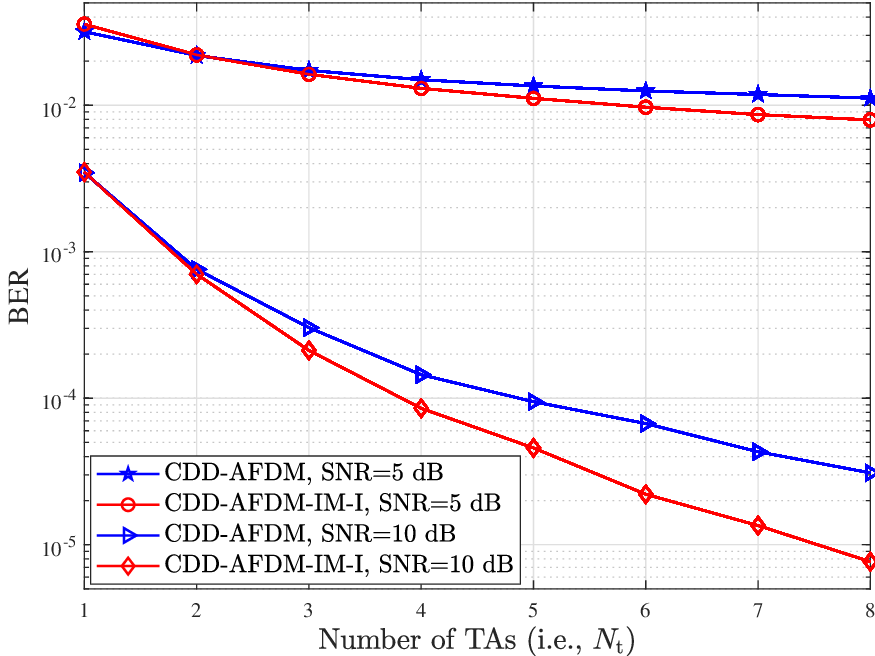}
	\vspace{-0.2cm}
	\caption{BER performance of the proposed CDD-AFDM-IM-I scheme and CDD-AFDM scheme by using DLMP detection versus ${N}_{\rm{t}}$ over a LTV channel, where ${\rm{SNR}}={5}~\rm{dB}$, ${10}~\rm{dB}$.}
	\label{NT}  
	\vspace{-2mm}}
\end{figure}
In this subsection, we further evaluate the BER performance of the proposed DLMP algorithm for practical large-dimensional signal detection in the proposed CDD-AFDM-IM systems.
Unless otherwise stated, the number of subsymbols in the DAF domain is set to $N=64$. The number of groups for CDD-AFDM-IM-I scheme is set to $g=16$, while the number of  subsymbol blocks and the number of groups for each subsymbol blocks are $L=8$ and $g=2$ for CDD-AFDM-IM-II scheme, respectively. Without loss of generality, the number of activation subsymbols for each group is set to $m=1$.
{For the DLMP detection algorithm, the damping factor is set to $\varpi=0.2$, and the number of maximum iterations is ${n}_{\rm iter}^{\rm max}=20$, after extensive experimentations as a compromise between convergence speed and accuracy.}

In Fig.~\ref{int-fra-MP-com-I}, we compare the BER performances of the proposed CDD-AFDM-IM-I, CDD-AFDM, AFDM and OTFS schemes by using the DLMP detection over the LTV channel.
We apply the QPSK mapping for the modulated bits in CDD-AFDM-IM-I scheme, and adopt the BPSK mapping in CDD-AFDM, AFDM and OTFS schemes, leading to the same SE (i.e., $1~\rm{bits/Hz}$) for different schemes.
From Figs.~\ref{int-MP-com} and~\ref{fra-MP-com}, one can observe that the CDD-AFDM-IM-I scheme achieves the better BER performance than the benchmark schemes in both the integer and fractional Doppler cases.
For instance, as shown in Fig.~\ref{int-MP-com}, when ${N}_{\rm{t}}=4$, the proposed CDD-AFDM-IM-I scheme yields an about $1.3~{\rm{dB}}$ gain compared to the CDD-AFDM scheme at a BER level of ${10}^{-4}$ with the integer Doppler.
Similarly for the fractional Doppler case, when ${N}_{\rm{t}}=5$, the proposed CDD-AFDM-IM-I scheme achieves an about $1.5~{\rm{dB}}$ gain with respect to the CDD-AFDM scheme at a BER level of ${10}^{-5}$.
Besides, one can notice that AFDM and OTFS can achieve similar performance over the LTV channel.
However, the one-dimensional representation of the delay-Doppler channel in AFDM allows a lower pilot overhead than OTFS.
This indicates that the AFDM scheme can be considered as a competitive LTV channel solution.

{In addition, we investigate the BER performance of the CDD-AFDM-IM-I scheme and CDD-AFDM versus the number of TAs in Fig.~\ref{NT}.
It can be observed from Fig.~\ref{NT} that the performance gain of the CDD-AFDM-IM-I scheme compared to the CDD-AFDM scheme gradually increases as the number of TAs increases.
This reason is that the paths between the TAs are independent and the total paths increase with the number of TAs, which results in a stronger intercarrier interference (ICI).
Compared with the CDD-AFDM scheme, the potential advantage of the proposed CDD-AFDM-IM-I scheme is that the subsymbols in the DAF domain are partially activated and the transmitted symbol in the DAF domain is characterized as a sparse vector.
This sparse frame structure can mitigate the impact of ICI on the BER performance of the CDD-AFDM-IM-I scheme.}
\begin{figure}[t]
        \centering
        \subfigure[\hspace{-0.2cm}]{\label{int-MP-com-II}
        \includegraphics[width=2.8in,height=2.0in]{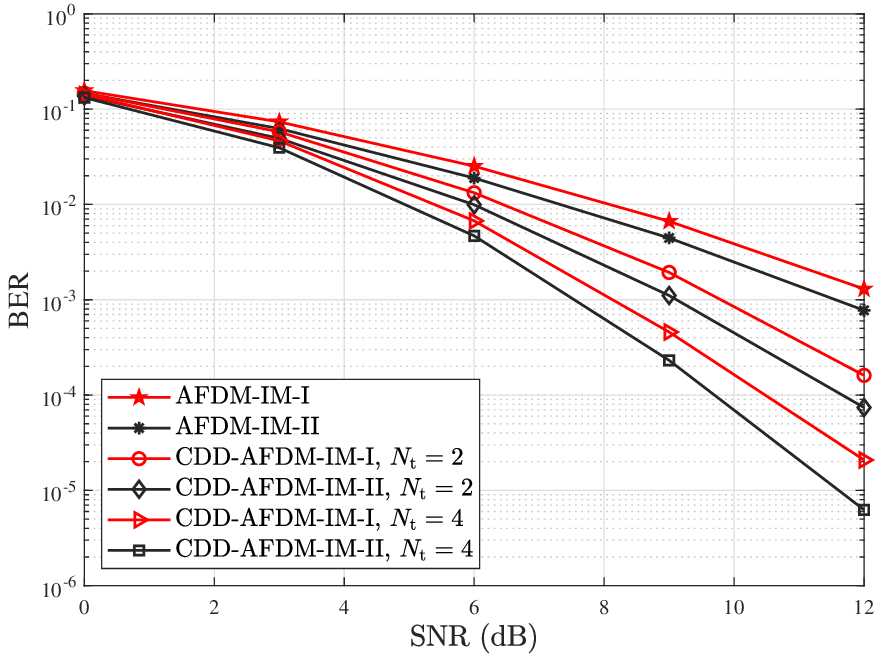}}
        \subfigure[\hspace{-0.2cm}]{\label{fra-MP-com-II}
        \includegraphics[width=2.8in,height=2.0in]{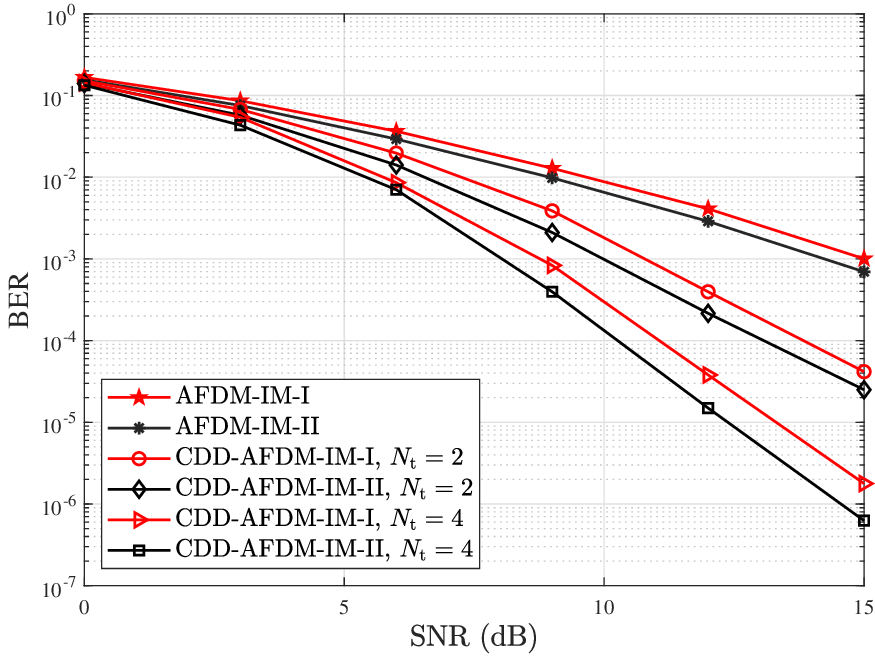}}
        \vspace{-0.3cm}
    	\caption{BER performance of the proposed CDD-AFDM-IM-II, CDD-AFDM-IM-I, AFDM-IM-II, and AFDM-IM-I schemes by using the DLMP detection for (a) integer Doppler and (b) fractional Doppler cases over an LTV channel.
    }\label{int-fra-MP-com-II}  
    	\vspace{-0mm}
\end{figure}

In Fig.~\ref{int-fra-MP-com-II}, we evaluate the BER performance of the proposed CDD-AFDM-IM-I, CDD-AFDM-IM-II, AFDM-IM-I and AFDM-IM-II schemes.
We apply the QPSK mapping for the modulated bits in CDD-AFDM-IM-II and AFDM-IM-II schemes, and adopt the BPSK mapping in CDD-AFDM-IM-I and AFDM-IM-I schemes, leading to the same SE as $0.75~\rm{bits/Hz}$ for different schemes.
One can notice that the BER performances of the CDD-AFDM-IM-II scheme and the AFDM-IM-II scheme outperform those of the CDD-AFDM-IM-I scheme and the AFDM-IM-I scheme for both integer and fractional Doppler cases over the LTV channel.
Specifically, when ${N}_{\rm{t}}=4$, the proposed CDD-AFDM-IM-II scheme achieves an around $1~{\rm{dB}}$ gain at a BER level of $10^{-4}$ than the CDD-AFDM-IM-I scheme for integer Doppler cases.
Similar performance gains can be observed in the fractional Doppler case.
These results strongly support the reliability and advantage of the CDD-AFDM-IM-II scheme over the CDD-AFDM-IM-I scheme for large-scale data~transmissions.

\begin{figure}[t]
        \centering
        \subfigure[\hspace{-0.5cm}]{\label{algo-com}
        \includegraphics[width=1.6in,height=2.0in]{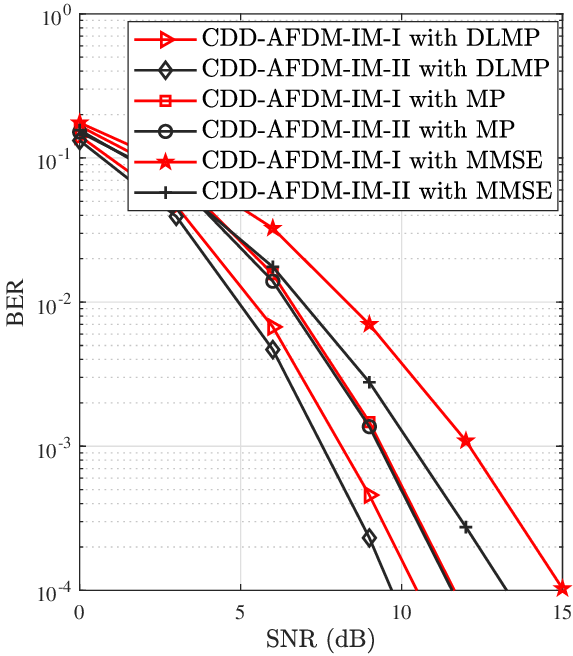}}
        \subfigure[\hspace{-0.4cm}]{\label{com-com}
        \includegraphics[width=1.6in,height=2.0in]{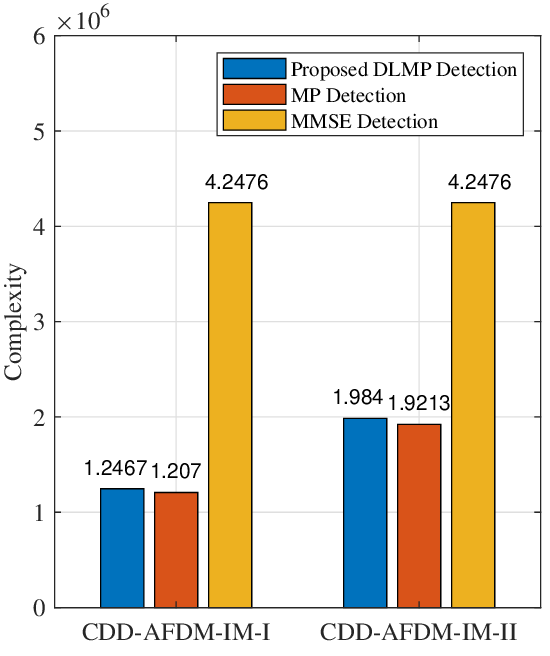}}
        \vspace{-0.3cm}
    	\caption{BER performance and computational complexity of the proposed DLMP, traditional MP and MMSE detectors for the large-scale CDD-AFDM-IM systems, (a) BER performance, (b) computational complexity (in terms of Flops), where the number of TAs is $4$ and the SE is $0.75$~\rm{bits/Hz}.}
    	\label{algo-com-com}  
    	\vspace{-0mm}
\end{figure}
To highlight the attractive tradeoff between the computation complexity and the BER performance of the proposed DLMP detector, we also provide the benchmark performance and complexity of traditional MMSE~\cite{tao2023affine} and single-layer MP~\cite{8424569} detectors in Fig.~\ref{algo-com-com}.
For clarity, we quantify the gains and losses in BER performance and computational complexity of different detection algorithms.
For example, consider the proposed CDD-AFDM-IM-I/II schemes with a target BER of $10^{-4}$ over an LTV channel.
For the CDD-AFDM-IM-I scheme, the MP and MMSE detections respectively require $31.83\%$ and $188.40\%$ higher SNRs compared to the proposed DLMP detection for the same BER target.{\footnote{Taking the CDD-AFDM-IM-I scheme as an example, the BER performance loss of MP detection with respect to DLMP detection can be calculated as $({10}^{\frac{{\rm SNR}_{\rm gap}}{10}}-1)\times{100\%}\approx{31.83\%}$, where ${\rm SNR}_{\rm gap}=1.2~{\rm dB}$ is the SNR gap to achieve the same BER target ($10^{-4}$ in this example).}}
Similarly, the MP and MMSE detections respectively require $58.49\%$ and $129.09\%$ higher SNRs than the proposed DLMP detection for the same BER target in the CDD-AFDM-IM-II scheme.
In terms of computational complexity, the complexity of the proposed DLMP detection is only $3.29\%$ higher than that of MP detection but $70.64\%$ lower than that of the MMSE detection for the CDD-AFDM-IM-I scheme.{\footnote{The increase or reduction computational complexity of the DLMP detection with respect to MP and MMSE detection can be calculated as $\lvert{\frac{{{\mathcal{O}}_{\rm{DLMP}}}}{{{\mathcal{O}}_{\rm{MP}}}}}-1 \rvert\times{100\%}\approx{3.29\%}$ and $\lvert{\frac{{{\mathcal{O}}_{\rm{DLMP}}}}{{{\mathcal{O}}_{\rm{MMSE}}}}}-1 \rvert\times{100\%}\approx{70.64\%}$, respectively.}}
Similarly, the complexity of the designed DLMP detection is only $3.26\%$ higher than that of MP detection but $53.29\%$ lower than that of the MMSE detection for the CDD-AFDM-IM-II scheme.
The above results show that for the CDD-AFDM-IM-I/II scheme, the proposed DLMP detection achieves a significant performance improvement compared to both MMSE and MP detections, and the complexity is slightly higher than that of MP detection but significantly lower than that of MMSE.
This verifies that the proposed DLMP detection has a better tradeoff between the complexity and the BER performance than the benchmark detections.

\begin{figure}[t]
	\center{
	\includegraphics[width=2.8in,height=2.0in]{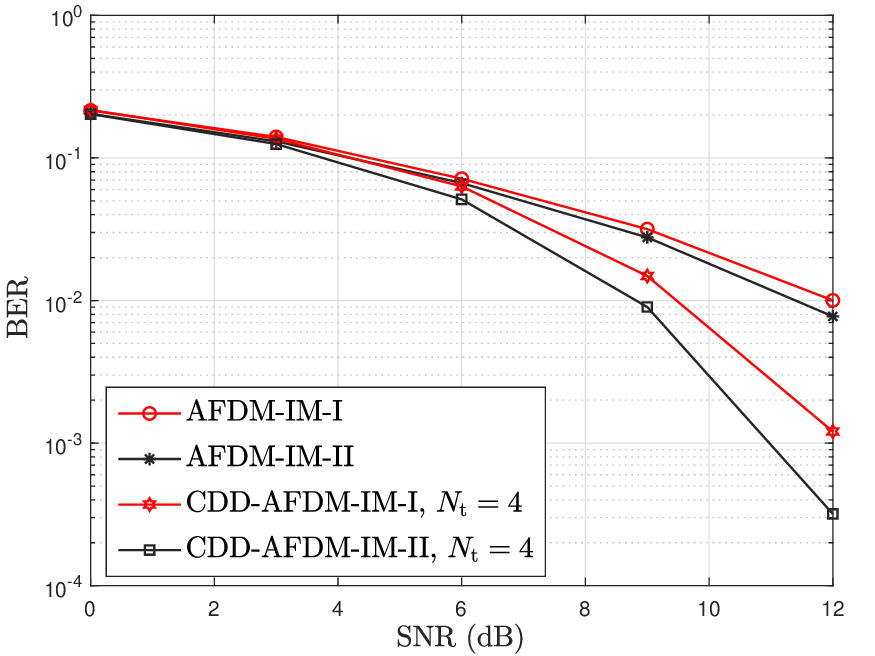}
	\vspace{+0.2cm}
	\caption{BER performance of the proposed CDD-AFDM-IM-I, CDD-AFDM-IM-II, AFDM-IM-I and AFDM-IM-II schemes by using the DLMP detection for the higher-order QAM over a LTV channel.}
	\label{RE3Q7_V1}  
	\vspace{-0mm}}
\end{figure}

In Fig.~\ref{RE3Q7_V1}, we further explore the performance of the proposed CDD-AFDM-IM schemes under higher-order QAM modulation, where the number of activator symbols per group is set to $m=2$ for the CDD-AFDM-IM-I and AFDM-IM-I schemes, and $m=3$ for the CDD-AFDM-IM-II and AFDM-IM-II schemes.
Also, we apply the $16$-QAM mapping for the modulated bits in the CDD-AFDM-IM-I and AFDM-IM-I schemes, and adopt $8$-QAM mapping in the CDD-AFDM-IM-II and AFDM-IM-II schemes, leading to the same SE (i.e., $2.5~{\rm bits/Hz}$) for different schemes. As shown in Fig.~\ref{RE3Q7_V1}, the proposed CDD-AFDM-IM-II and AFDM-IM-II schemes outperform the proposed CDD-AFDM-IM-I and AFDM-IM-I schemes, respectively. These results once again demonstrate the reliability and advantages of the CDD-AFDM-IM-II scheme over the CDD-AFDM-IM-I scheme in terms of large-scale antennas as well as high spectral efficiency transmissions.

\section{Conclusion}
\label{section:Conclusion}
In this paper, we proposed two typical CDD-AFDM-IM schemes for high-mobility communications.
We have derived the full diversity conditions of the proposed CDD-AFDM-IM schemes for both the integer and fractional Doppler cases over the LTV channel.
Besides, we presented the design guidelines for the high-reliability IM scheme under the multiple-antenna AFDM transmission framework from the diversity perspective.
The theoretical BER upper bounds of the proposed schemes with ML detection were analyzed and verified by the simulation.
We also proposed a low-complexity DLMP algorithm for practical large-dimensional signal detection of the proposed CDD-AFDM-IM schemes.
The results demonstrated that the proposed DLMP detection achieves a better tradeoff between computational complexity and BER performance.
Finally, the performance comparison with the benchmark schemes illustrated that our proposed CDD-AFDM-IM schemes achieve excellent performance over the LTV channel.
{In summary, the proposed CDD-AFDM-IM schemes represent a competitive solution for future high-mobility communications. In the future, we will address the security of AFDM schemes by integrating encryption, secure key management, and adaptive mechanisms to combat jamming and spoofing, thereby improving their suitability for secure communications. Also, we will consider the design of AFDM schemes in multi-user scenarios to provide a more comprehensive assessment of system performance.}

\bibliographystyle{IEEEtran}  

\end{document}